\documentclass[11pt]{article}

\usepackage{arxiv}
\usepackage[utf8]{inputenc} 
\usepackage[T1]{fontenc}    
\usepackage{hyperref}       
\usepackage{url}            
\usepackage{booktabs}       
\usepackage{amsfonts}       
\usepackage{nicefrac}       
\usepackage{microtype}      
\usepackage{nomencl}
\usepackage{amssymb}
\usepackage{amsmath}
\usepackage[noabbrev,capitalise,sort&compress]{cleveref}  
\usepackage{subcaption}
\usepackage[normalem]{ulem}
\hypersetup{colorlinks,urlcolor=blue}
\usepackage[bbgreekl]{mathbbol}
\usepackage{etoolbox}
\usepackage{mathtools}
\usepackage{placeins}
\usepackage{algorithm}
\usepackage{algpseudocode}
\usepackage{pgfplots}
\usepgfplotslibrary{groupplots}
\pgfplotsset{compat=newest}
\usepackage[style=numeric-comp,backend=biber, giveninits=true]{biblatex}
\DeclareSymbolFontAlphabet{\mathbb}{AMSb}
\DeclareSymbolFontAlphabet{\mathbbl}{bbold}

\newcommand{\yobs}{{\boldsymbol{y}_{\text{obs}}}}

\newcommand{\DM}{\mathfrak{M}}

\newcommand{\like}{\pdf{\yobs|\lv}}
\newcommand{\prior}{\pdf{\lv}}
\newcommand{\posterior}{\pdf{\lv|\yobs}}
\newcommand{\evidence}{\pdf{\yobs}}
\newcommand{\pdf}[1]{p\left(#1\right)}
\newcommand{\unpostf}{\pi\left(\yobs,\lv\right)}
\newcommand{\unpostfsurrogate}{\pi^\text{s}\left(\yobs,\lv\right)}
\newcommand{\unpostfs}{\pi\left(\yobs,\lvs\right)}

\newcommand{\Ex}[2]{{\mathbb{E}}_{#1}\left[#2\right]}
\newcommand{\Var}[2]{\mathbb{V}_{#1}\left[#2\right]}

\newcommand{\Cov}[2]{\mathbbl{Cov}_{#1}\left[#2\right]}
\newcommand{\cov}[2]{\mathbbl{COV}_{#1}\left[#2\right]}
\newcommand{\nd}[2]{\mathcal{N}_{#1}\left(#2\right)}

\newcommand{\norm}[1]{\left\lVert#1\right\rVert}

\newcommand{\sample}[2]{#1^{(#2)}}
\newcommand{\sams}[1]{\sample{#1}{s}} 
\newcommand{\lv}{\boldsymbol{\theta}}
\newcommand{\lvj}{\theta_j}

\newcommand{\lvs}{\sams{\lv}}
\newcommand{\isws}{\sams{w}}
\newcommand{\varp}{\boldsymbol{\lambda}}
\renewcommand{\varpi}{\varp^{i}}
\newcommand{\varq}[1]{q(#1)}

\newcommand{\varqf}{\varq{\lv|\varp}}
\newcommand{\varqfi}{\varq{\lv|\varpi}}
\newcommand{\varqisf}{q_\text{is}^i\left(\lv\right)}
\newcommand{\varqisfs}{q_\text{is}^i\left(\lvs\right)}
\newcommand{\varqisw}{\beta_j^i}
\newcommand{\score}[2]{\grad{\ln #2}{#1}}
\newcommand{\scoreq}{\score{\varp}{\varqf}}
\newcommand{\scoreqi}{\score{\varp}{\varqfi}}

\newcommand{\ELBO}{\mathcal{L}}

\newcommand{\gradscore}[2]{\nabla^\text{sc}_{#2}#1}
\newcommand{\grad}[2]{\nabla_{#2}#1}

\newcommand{\niter}{N_\text{it}}
\newcommand{\memseti}{\mathcal{Q}^i}
\newcommand{\ess}{\text{ESS}}
\newcommand{\matrixnorm}[2]{\norm{#1}_{#2}}

\newcommand{\card}[1]{\text{card}(#1)}

\newcommand{\sseti}{\Theta^i}
\newcommand{\mseti}{\Phi^i}
\newcommand{\pseti}{\Lambda^i}
\newcommand{\ssetj}{\Xi^j}
\newcommand{\pfrac}[2]{\frac{\partial #1}{\partial #2}}

\newcommand{\vel}{\dot{\boldsymbol{u}}_\text{def}}
\newcommand{\acc}{\ddot{\boldsymbol{u}}_\text{def}}

\newlength{\commentindent}

\allowdisplaybreaks 
\title{A Black Box Variational Inference Scheme for Inverse Problems with Demanding Physics-Based Models}

\author{
  *Gil Robalo Rei\\
  Institute for Computational Mechanics\\
  Technical University of Munich\\
  D-85748 Garching b. München\\
  \texttt{gil.rei@tum.de} \\
  \And
  Christoph P. Schmidt\\
  Institute for Computational Mechanics\\
  Technical University of Munich\\
 \And
  Jonas Nitzler\\
  Institute for Computational Mechanics\\
  Technical University of Munich
  \And
  Maximilian Dinkel\\
  Institute for Computational Mechanics\\
  Technical University of Munich\\
  \And
Wolfgang A. Wall\\
Institute for Computational Mechanics\\
  Technical University of Munich\\
}

\begin{document}
\maketitle

\begin{abstract}
Bayesian methods are particularly effective for addressing inverse problems due to their ability to manage uncertainties inherent in the inference process. However, employing these methods with costly forward models poses significant challenges, especially in the context of non-differentiable models, where the absence of likelihood model gradient information can result in high computational costs. To tackle this issue, we develop a novel Bayesian inference approach based on black box variational inference, utilizing importance sampling to reuse existing simulation model calls in the variational objective gradient estimation, without relying on forward model gradients. The novelty lies in a new batch-sequential sampling procedure, which only requires new model evaluations if the currently available model evaluations fail to yield a suitable approximation of the objective gradient. The resulting approach reduces computational costs by leading to variational parameter updates without requiring new model evaluations when possible, while adaptively increasing the number of model calls per iteration as needed. In combination with its black box nature, this new approach is suitable for inverse problems involving demanding physics-based models that lack model gradients. We demonstrate the efficiency gains of the proposed method compared to its baseline version, sequential Monte Carlo, and Markov-Chain Monte Carlo in diverse benchmarks, ranging from density matching to the Bayesian calibration of a nonlinear electro-chemo-mechanical model for solid-state batteries.

\end{abstract}

\newpage
\section{Introduction}
Physics-based computational models have become an essential tool to predict real-world phenomena in numerous  fields of engineering and the applied sciences \cite{Schmidt2023,Brandstaeter2018}. To make meaningful predictions, problem-specific parameters, e.g.~material parameters or boundary conditions, need to be calibrated based on experimental data, oftentimes corrupted by noise. Even for uncorrupted data, classical deterministic inference approaches such as regularized least-squares, not only lead to ill-posed problems but also only infer a single parameter set. It is unclear whether this parameter set genuinely reflects the optimal result from the measured data or if it is simply a random outcome shaped more by the numerical properties of the optimization method than by meaningful insights from the specific experiments and measurements. To handle this ill-posedness, as well as to intrinsically incorporate uncertainty in the calibration process, Bayesian approaches have emerged \cite{Willmann2022,Hervas-Raluy2023, Bilionis2014}. Here, using Bayes' theorem, the inference task is reformulated into a Bayesian inverse problem (BIP):
\begin{align}
    \posterior = \frac{\like \prior}{\evidence},
\end{align}
where $\prior$ is the prior distribution on the parameters of interest $\lv$, also called latent variables, $\like$ the likelihood linking the forward model $\DM(\lv)$ to the observed data $\yobs$ and $\posterior$ the posterior distribution of the latent variables. Consequently, evaluating the likelihood involves computing the forward model $\DM(\lv)$, which for computationally expensive models is a resource-intensive operation. The evidence $\evidence$ represents a normalization constant between the probabilistic model $\unpostf=\like\prior$ and the posterior distribution. In Bayesian inverse analysis, the goal is to infer the posterior distribution of the latent parameters. This approach to inverse problems is generally well-posed \cite{Stuart2010}, as the inferred quantity assesses the entire parameter space, also called sample space. This can only be done analytically for specific probabilistic models. Therefore, in general, inference schemes solely relying on model interrogations to explore the sample space are employed instead.

Most inference schemes can be categorized into sampling-based, particle-based, or variational inference approaches. The most prominent sampling-based inference method is Markov chain Monte Carlo (MCMC), which approximates the posterior distribution by generating sample sets of the true posterior distribution. Depending on the problem setting, this approach can require millions of probabilistic model calls to construct accurate approximations \cite{Bilionis2014}. To reduce inference time, a second class of inference schemes based on particle filters was proposed. Here, the posterior is approximated as a mixture of particles which, based on a transition kernel, are iteratively moved within the sample space to regions of high probability mass \cite{Chopin2002, DelMoral2006} with a certain probability. One notable candidate of this approach is sequential Monte Carlo (SMC)  \cite{Chopin2002, DelMoral2006, Dau2022}. The last main category is variational inference (VI) \cite{Jordan1998, Hoffman2013}, for which the posterior distribution type is preselected, and its parameterization is obtained via the minimization of a discrepancy between the assumed and true posterior distribution. Due to the choice of posterior approximation, the computational cost of the inference process can be drastically reduced compared to sampling-based approaches \cite{Bilionis2014}.

Independent of the inference method, approximating an intractable posterior can still require thousands of probabilistic model calls. For computationally costly forward models $\DM$, it is vital to keep the number of forward model calls as small as possible; otherwise, the inference process would be prohibitively expensive. For many real-world problems in engineering and beyond, sophisticated numerical models are needed to predict elaborate physical system behavior. In particular, we focus on black box approaches capable of addressing situations where model derivatives are either difficult to obtain or unavailable, which is often the case with proprietary software tools. Such a setting does not require insights into the inner workings of the model and can, therefore, be applied to a large range of applications. From a performance perspective, gradient-based variational inference approaches~\cite{Bilionis2014} have demonstrated significant potential for efficient inference in terms of model evaluations. To leverage the advantage of such approaches in a black box setting, we build upon black box variational inference~\cite{Ranganath2014}, exploiting importance sampling (IS) as a means to reduce the number of model calls, resulting in a flexible yet efficient approach suited for expensive physics-based forward models, especially those in which model derivatives are unavailable.

In the next section, we highlight the difficulties and desired properties for an inference scheme in combination with expensive forward models. The remainder of the paper consists of insights into the proposed inference algorithm and, finally, numerical examples.

\section{Inference with Demanding Black Box Physics-Based Models}
This work focuses on demanding physics-based simulation models. These are often the outcome of decades of expertise in modeling complex systems, combined with high-performance implementations to exploit available hardware resources to predict real-world scenarios. Consequently, the codes are purpose-built for their applications without the necessity for efficient inverse analysis in mind. To flexibly infer the posterior distribution independent of the underlying model, a black box approach is vital. If the model is differentiable, gradient-based inference schemes could speed up the inference time drastically \cite{Bilionis2014}. However, obtaining model derivatives is often impractical due to the legacy nature of the codes. Automatic differentiation (AD) requires using particular packages that are not necessarily compatible with the existing code base regarding infrastructure or performance. A valid alternative is using the adjoint method \cite{Bilionis2014}; however, this can require additional problem-specific modeling to construct the model-dependent adjoint problem.
In contrast, obtaining the derivatives by computing finite differences (FD) is straightforward. However, depending on the dimension of $\lv$, constructing the FD stencil requires multiple model evaluations to compute a single derivative, increasing the computational costs drastically and eventually outgaining the advantage of the derivative information in the inference process. Given these conditions, employing gradient-based inference schemes is typically impractical. Hence, approaches without differentiability requirements on the probabilistic model are needed. Here, no local sensitivities can be exploited, such that most black box approaches encounter difficulties in the case of sparse posteriors relative to their sample space. Since sparsity is more prevalent in high-dimensional sample spaces, black box inference methods become even less efficient than their gradient-based counterparts in higher dimensions. In this work, we limit ourselves to 32-dimensional cases, although no specific limit exists.

To diminish the computational cost associated with the model wall times, a natural approach to speed up the inference process is to replace the probabilistic model with a surrogate $\unpostfsurrogate\approx\unpostf$ that is cheaper to evaluate. Commonly, regression approaches are used to construct an approximation based on true model queries. As long as the training process requires fewer true model calls than direct inference would, a substantial amount of computational budget is saved. Different approaches exist, e.g., from replacing the forward model \cite{Hervas-Raluy2023} to replacing the entire probabilistic model \cite{Willmann2022, Dinkel2024}. If multiple models of various accuracies are available, employing multi-fidelity methods \cite{nitzler2025, Koutsourelakis2009a, Perdikaris2017, Nitzler2022} have shown accurate results on a low-cost computational budget. However, when employing surrogate approaches, it is crucial to capture essential features of the true model, such as length scales or smoothness properties, thus requiring additional careful problem-specific modeling. Constructing accurate surrogate models might be prohibitively expensive depending on the available knowledge of model behavior and model in/output dimensions. For these cases, we require an efficient direct inference approach where only a small number of model calls are needed.

Nearly all inference approaches can be cast into a batch sequential framework, i.e., its posterior approximation is updated iteratively based on a batch of $n_\text{batch}$ independent probabilistic model queries. When working with expensive probabilistic models, the batch evaluations dominate inference time. In computational engineering, the models tend to be implemented in a high-performance environment, generally able to distribute the computation over multiple processing units. Since the batch evaluation can be done embarrassingly parallel, this leads to a nested parallel setting, i.e., multiple simulations, each distributed over multiple CPUs. Therefore, parallelizable inference approaches are crucial to exploit existing computational resources to reduce the time to solution while minimizing the total number of model calls required.

\section{Adaptive Batch-Sized Regulated Importance Sampling based Black Box Variational Inference}

As mentioned, in this work, we build upon black box variational inference (BBVI) \cite{Ranganath2014} to develop a variational inference approach tailored to Bayesian inverse problems with expensive models. The main idea of variational inference is to transform the inference task into an optimization problem \cite{Jordan1998, Hoffman2013}. For this, a variational distribution family $\varqf$ is chosen to approximate the exact posterior, and a divergence measure between the densities is minimized. However, as the exact posterior is unknown, for the reverse Kullback-Leibler divergence, the minimization can be reformulated using the evidence lower bound (ELBO) \cite{Blei2017}
\begin{align}\label{eq:elbo_max}
\varp^\text{opt}&\in \underset{\varp}{\max} \ \underbrace{\Ex{\varqf}{\ln \unpostf - \ln \varqf}}_{\ELBO(\varp)},
\end{align}
where $\Ex{\varqf}{\cdot}$ is the expectation operator w.r.t. $\varp$. The ELBO $\ELBO(\varp)$ is a lower bound on the log evidence (hence the name), and for a given probabilistic model and variational distribution solely depends on the variational parameters $\varp$. Stochastic variational inference \cite{Hoffman2013, Zhang2017} iteratively updates the variational parameters by
\begin{align}\label{eq:stochastic_optimizer}
    \varp^{i+1}=\varpi+\boldsymbol \alpha ^{i}\nabla_{\varp}\ELBO(\varpi).
\end{align}
The learning rate $\boldsymbol \alpha^{i}$ is commonly set based on the Robbin-Monroe conditions \cite{Robbins1951, Ranganath2014} or stochastic optimizers \cite{Kingma2014, dinkel2024dlrd}. In this notation, the learning rate implicitly includes gradient preconditioning. For example, $\boldsymbol \alpha^{i}$ might include the inverse Fisher information matrix (FIM) of $\varqf$ \cite{Fisher1921, Amari1998} which is an approximation to the ELBO-Hessian \cite{Tang2019}. Insights and limitations of this approach can be found in \cite{Martens2020, Tang2019} and \cite{Kunstner2019}, respectively. The missing ingredient in \eqref{eq:stochastic_optimizer} is the ELBO gradient w.r.t.~the variational parameters. The expectation of the ELBO gradient is oftentimes intractable, such that it is commonly approximated with Monte Carlo (MC) gradients \cite{Mohamed2020}.  As the focus of this work lies in black box problems without access to the gradient of the probabilistic model, the gradient in \eqref{eq:stochastic_optimizer} needs to be approximated using black box estimators (w.r.t.~the probabilistic model). For this, at iteration $i$, based on the score function $\scoreqi=\scoreq\big|_{\varp=\varpi}$, the score function estimator \cite{Williams1992,Rubinstein1986,Ranganath2014,Mohamed2020}
\begin{align}\label{eq:score_function_estimator}
    \gradscore{\ELBO}{\varp}(\varpi)&=\Ex{\varqfi}{\underbrace{\scoreqi\left(\ln \unpostf - \ln \varqfi\right)}_{\boldsymbol{g}^\text{sc}(\lv,\varpi)}}\\
    &\approx\frac{1}{N}\sum_{s=1}^N \score{\varp}{\varq{\lvs|\varpi}}\left(\ln \unpostfs - \ln \varq{\lvs,\varpi}\right) 
\end{align}
is employed, where a batch of $N$ samples $\lvs$ is drawn from $\varqfi$. As can be seen, no derivatives of the probabilistic model are needed, fulfilling the desired black box property. The variance of this estimator is generally large, particularly compared to model-gradient-based estimators \cite{Mohamed2020}, such that variance reduction techniques are required \cite{Ranganath2014, Kool2019, Uchibe2018, Ruiz2016}. In this work, to reduce estimator variance, we employ a baseline estimator \cite{Ranganath2014, Mohamed2020}
\begin{equation}\label{eq:baseline_estiamtor}
    \boldsymbol{g}(\lv,\varpi)=\boldsymbol{g}^\text{sc}(\lv,\varpi) - a\scoreqi,
\end{equation}
where the score function is used as a control variate \cite{Ranganath2014}. The coefficient $a$ regulates the variance reduction effect. The score function is an interesting choice as a control variate for two reasons. First, its expectation is known analytically, independent of the distribution \cite{Ranganath2014}
\begin{equation}\label{eq:score_function_property}
   \Ex{\varqf}{\scoreq}=\boldsymbol{0}.  
\end{equation}
Secondly, the component $c$ of the ELBO gradient can be interpreted as the covariance between the score function and inner cost function  $\ln \unpostf - \ln \varqfi$ \cite{Mohamed2020}
\begin{equation}\label{eq:elbo_score_correlation}
    \gradscore{\ELBO}{\lambda_c}(\varpi)=\cov{\varqfi}{\score{\lambda_c}{\varqfi}, \ln \unpostf - \ln \varqfi},
\end{equation}
indicating a strong correlation between $\boldsymbol{g}^\text{sc}$ and $\scoreq$. Since the score function is not dependent on the model, it is cheap to evaluate. 
Even with variance reduction, variational inference may still require thousands of iterations $\niter$ to reach convergence. For real-world engineering problems, due to the expensive models, the number of necessary model calls is proportional to the number of Monte Carlo samples $N$ in \eqref{eq:score_function_estimator}, leading to enormous computational costs and making this approach impractical. Since the expectation in \eqref{eq:stochastic_optimizer} is w.r.t.~the variational distribution of the current iteration $\varqfi$, previous evaluations $j<i$ can not be directly reused as they were drawn from a different distribution $\varq{\lv|\varp^{j}} \neq \varqfi$. To incorporate these samples in the current estimation without introducing a bias, an importance sampling (IS) approach is applied
\begin{align}
    \gradscore{\ELBO}{\varp}(\varpi)&=\Ex{\varqfi}{\boldsymbol{g}(\lv,\varpi)}=\Ex{\varqisf}{w(\lv)\boldsymbol{g}(\lv,\varpi)}\label{eq:is_elbograd} \\
    &\approx \frac{1}{N}\sum_{s=1}^N \underbrace{\frac{\varq{\lvs|\varpi}}{\varqisfs}}_{\isws}\boldsymbol{g}\big(\lvs,\varpi\big)\label{eq:is_elbograd_MC},  
\end{align}
where the weights $\isws$ are the relative likelihood ratio between variational and importance sampling proposal distribution $\varqisf$ from which the samples $\lvs$ are drawn. The proposal distribution is chosen as
\begin{align}
    \varqisf = \sum_{j\ \in \memseti}\varqisw \varq{\lv|\varp^j}\label{eq:is_distribution},
\end{align}
where $\memseti$ is a set of indices related to variational parameters of previous iterations where new samples were drawn. Hence, the importance sampling distribution is a mixture model with components consisting of the variational distribution at previous iterations. Drawing samples of this mixture model is a two-step process:
\begin{enumerate}
    \item Sample a component $\sample{\gamma}{l}$ from a discrete distribution with probability $P(\gamma=j)=\varqisw$ for $j\ \in \memseti$
    \item Draw a sample of this component $\sample{\lv}{l}\sim\varq{\lv|\varp^{\sample{\gamma}{l}}}$ 
\end{enumerate}
Constructing the proposal distribution as a mixture of previous distributions allows for exploitation in the second step to reuse $\lvs$ of earlier iterations, as the probabilistic model has already been evaluated for these samples. The enhanced number of samples in \eqref{eq:is_elbograd_MC} leads to variance reduction in the gradient estimate, decreasing the required iterations for convergence.
For this, the previously evaluated samples $\sample{\lv}{l} \in \ssetj$ in iteration $j$ are grouped into a sample set $\sseti$. The corresponding probabilistic model calls and variational parameters are stored in $\mseti$ and $\pseti$, respectively. The cardinality $\card{\memseti}$ defines the number of components of the IS proposal distribution. A high-level overview of an importance-sampling-based VI algorithm to reuse model calls can be found in \cref{alg:is_BBVI}.

\begin{algorithm}[H]
\caption{General importance-sampling-enhanced BBVI to reuse sample evaluations}\label{alg:is_BBVI}
\begin{algorithmic}
\State $\Theta^{0},\ \Phi^{0},\ \Lambda^{0}=\{\emptyset\}$
\While{not converged}
    \State $\sseti$, $\mseti$, $\pseti$ $\gets$ \texttt{update\_sets($\varpi$, $\sseti$, $\mseti$, $\pseti$)} \Comment{Update sets, e.g. algorithm \ref{alg:abris_sampling_loop}}
    \State $\varqisw =$ \texttt{compute\_mixture\_coefficients($\varpi$, $\sseti$, $\mseti$, $\pseti$)} \Comment{Compute mixture coefficients}
    \State $\isws=\frac{\varq{\lvs|\varpi}}{\varqisfs}$, $\ \forall \lvs \in \sseti$ \Comment{Compute IS sample weights}
    \State $\gradscore{\ELBO(\varpi)}{\varp} \approx \frac{1}{M}\sum_{s=1}^M \isws\boldsymbol{g}\big(\lvs,\varpi\big)$ \Comment{IS-based MC}
    \State $\varpi\gets\varpi+\boldsymbol \alpha^{i} \gradscore{\ELBO(\varpi)}{\varp}$ \Comment{Update the variational parameters}
    \State $i \gets i+1$
\EndWhile
\end{algorithmic}
\end{algorithm}

To complete the importance-sampling-based black box variational inference approach, an update procedure for sample sets $\sseti$, $\mseti$, $\pseti$ and mixture weights $\varqisw$. The update procedure has two main functions: requiring model calls and sample selection. The former dictates when new model evaluations are necessary. The simplest approaches consist of sampling the forward model at every iteration \cite{Uchibe2018, Sakaya2017, Arenz2018, Arenz2020}. Here, as for variance reduction approaches for \eqref{eq:score_function_estimator}, the computational gain stems from a reduced number of required iterations due to increased gradient accuracy. Yet, such an approach might still need many model calls. To make IS-VI applicable to expensive forward models, it becomes essential to relax the requirement of evaluating the model at each iteration. If no model calls are done within the current iterations, the variational parameters are updated solely based on this information. Considering expensive models, the posterior updates are virtually free regarding computational costs, leading to massive speed-ups compare to its base algorithm. Although the probabilistic model is not newly evaluated in these iterations, the gradient is recomputed at the current location, leading to nonlinear parameter updates w.r.t. $\varpi$.

If no new model evaluations were added, the parameter space would not be explored, leading to biased posterior approximations. To reduce this bias, a sampling strategy is required, which dictates when the model needs to be evaluated. This strategy dictates the construction of $\memseti$ and therefore $\sseti$, $\mseti$ and $\pseti$. Ideally, new samples for the gradient estimation should only be required if $\sseti$ and $\mseti$,i.e., the existing probabilistic model call evaluations, cannot produce a suitable approximation. To determine the quality of the sample set, Arenz et al.~\cite{Arenz2020} used the effective sample size (ESS), which represents the number of samples necessary for a direct Monte Carlo estimator to produce an IS-based estimate with the same variance \cite{Kong1994,Elvira2022}. Since it can not generally be computed in closed form, the ESS is commonly approximated as 
\begin{align}
    \ess(\sseti, \pseti, \varpi)=\frac{\left(\sum\limits_{s=1}^M \isws\right)^2}{\sum\limits_{s=1}^M \left(\isws\right)^2},
\end{align}
where $M$ is the number of previously evaluated samples in $\sseti$. Hence, $\ess(\sseti, \varpi)\leq N$ indicates that the sample sets are deteriorated such that $\sseti$ requires to be regenerated with new samples. A similar criterion has shown interesting results in \cite{Arenz2020}; however, using it as the sole condition has one disadvantage. Since the ESS solely depends on the existing sample sets and the importance sampling distribution, this criterion is not directly related to the ELBO, its gradient,  or variational inference. To overcome this shortcoming, we add a second criterion tailored to BBVI following the argumentation of the baseline estimator \eqref{eq:baseline_estiamtor} exploiting the correlation between ELBO gradient and score function \eqref{eq:elbo_score_correlation}. Due to the known expectation of the score function \eqref{eq:score_function_property}, the errors of its MC estimates can be computed at each iteration $i$ by
\begin{align}
   \boldsymbol{e}_\text{is}(\sseti, \pseti, \varpi)&=\frac{1}{M}\sum_{s=1}^M \score{\varp}{\varq{\lvs|\varpi}}\isws\\
   \boldsymbol{e}_\text{ref}(\varpi)&=\frac{1}{N}\sum_{s=1}^N \score{\varp}{\varq{\sample{\lv}{s}|\varpi}},
\end{align}
where $\boldsymbol{e}_\text{is}$ is computed using importance sampling with $\sseti$. The estimate $\boldsymbol{e}_\text{ref}$ is obtained using new samples $\sample{\lv}{l}$ of $\varqfi$. If the error of the estimate obtained via importance sampling is larger than the reference estimate, e.g. using an inverse $\boldsymbol{A}$-matrix norm $\matrixnorm{\boldsymbol{e}_\text{is}(\sseti, \pseti, \varpi)}{\boldsymbol{A}}>\alpha_\text{sc}\matrixnorm{\boldsymbol{e}_\text{ref}(\varpi)}{\boldsymbol{A}}$, new model calls will be triggered. Due to the draw of new samples $\sample{\lv}{l}$, the reference estimate is inherently stochastic without requiring probabilistic model evaluations. Since the samples are freshly drawn at every iteration, the probability of sampling previously unseen regions is increased. If these regions of the sample space are relevant for the estimation, the error in reference value will be smaller, such that new model calls will be requested to increase the exploration rate. The scaling factor $\alpha_\text{sc}$ depends on the choice of variational distribution. For unimodal variational distributions, we set $\alpha_\text{sc}=1$. For more expressive distributions, it is beneficial to increase this parameter, to avoid unnecessary draw of samples, particularly in the beginning of the inference process, where the varitional distribution can change radically.

Sakaya et al. \cite{Sakaya2017} proposed an approach to draw new samples within an iteration $i$ if a uniform random variable sample exceeds a predefined threshold independent of the current sample set. In contrast to the previous conditions, it is independent of the existing sample sets and focuses on the number of posterior updates. We leverage a similar idea by prescribing a periodicity of $N_\text{periodic}$ that triggers new model calls every $N_\text{periodic}$ iterations and acts as a safety layer to enforce exploration.

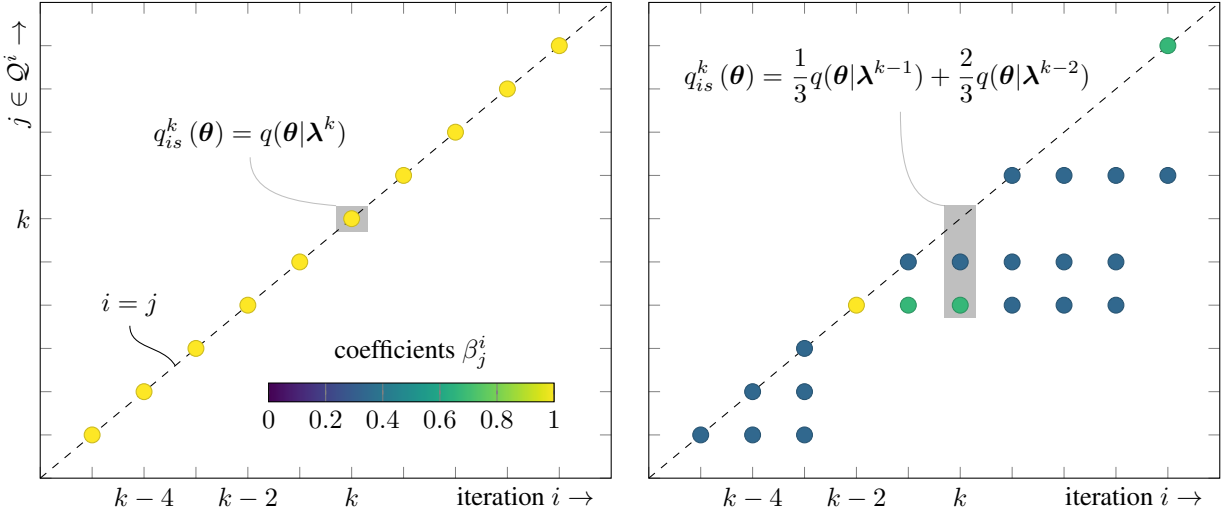
\begin{figure}
    \centering
    \newcommand{\varqisfk}{q_{is}^k\left(\lv\right)}

\begin{tikzpicture}
  \begin{groupplot}[
    group style={
      group size=2 by 1, 
      horizontal sep=0.5cm, 
    },
    colormap/viridis,
    width=\textwidth/1.8
  ]
   \nextgroupplot[
    x label style={at={(axis description cs:0.85,0)},anchor=north},
    y label style={at={(axis description cs:0,0.85)},anchor=south},
    xlabel={iteration $i \rightarrow$},
    ylabel={$j \in \memseti \rightarrow$},
    xmin=0, xmax=11,
    ymin=0, ymax=11,
    xtick={0,1,2,3,4,5,6,7,8,9,10},
    xticklabels={,,$k-4$,,$k-2$,,$k$,,,,},
    ytick={0,1,2,3,4,5,6,7,8,9,10},
    yticklabels={,,,,,,$k$,,,,},
    colorbar horizontal,
    colorbar style={
    at={(0.65,0.2)},anchor=north, 
    title=coefficients $\beta^i_j$,title style={at={(axis description cs:0.5,0.45)}, anchor=south},scale=0.5}
    ]
  ]
  \node[anchor=west] (varqis) at (axis cs:2,8){$\varqisfk=\varq{\lv|\varp^k}$};
  \node[anchor=west] (source_k) at (axis cs:5.7,6.3){};

  \draw[draw=gray, opacity=0.5] (varqis.south) to [out=-90,in=175] (source_k.west); 

  \node[anchor=west] (source) at (axis cs:1,4){$i=j$};
  \node (destination) at (axis cs:2.6,2.6){};
  \draw[] (source.south) to [out=-90,in=120] (destination.center);
  
  \addplot[gray, fill=gray, fill opacity=0.5,draw=none] coordinates {(5.7,5.7) (5.7,6.3) (6.3,6.3) (6.3,5.7)};

  \addplot[
    scatter,
    only marks,
    point meta=\thisrow{value},
    colormap/viridis, mark size = 3pt
  ] table [x=x, y=y, col sep=comma] {bbvi.csv};

  \draw[dashed] (axis cs: 0,0) -- (axis cs: 11,11);

%
%
%
%
%

  \nextgroupplot[
    x label style={at={(axis description cs:0.85,0)},anchor=north},
    xlabel={iteration $i \rightarrow$},
    xmin=0, xmax=11,
    ymin=0, ymax=11,
    xtick={0,1,2,3,4,5,6,7,8,9,10},
    xticklabels={,,$k-4$,,$k-2$,,$k$,,,,},
    ytick={0,1,2,3,4,5,6,7,8,9,10},
    yticklabels={,,,,,,,,,,},
  ]
    \node[anchor=west] (varqis) at (axis cs:0.5,9.3){$\begin{aligned}\varqisfk&=\frac{1}{3}\varq{\lv|\varp^{k-1}} +\frac{2}{3}\varq{\lv|\varp^{k-2}}\end{aligned}$};
    \node[anchor=west] (source_k) at (axis cs:5.7,6.3){};
  
    \draw[draw=gray, opacity=0.5] (varqis.south) to [out=-90,in=175] (source_k.west);

    \addplot[gray, fill=gray, fill opacity=0.5,draw=none] coordinates {(5.7,
  3.7) (5.7,6.3) (6.3,6.3) (6.3,3.7)};

    \addplot[
      scatter,
      only marks,
      point meta=\thisrow{value}, mark size = 3pt
    ] table [x=x, y=y, col sep=comma] {abris.csv};

    \draw[dashed] (axis cs: 0,0) -- (axis cs: 11,11);
  
    \end{groupplot}

\end{tikzpicture}
    \caption{Visualization of the sampling of BBVI (left) and its IS-enhancement ABRIS (right). The x-axis is the index of the stochastic optimizer iteration. The y-axis describes the iterations associated with samples in $\sseti$ used in the IS approach. Points on $i=j$ indicate that the model is newly sampled at this iteration, and off-diagonal points indicate that it has been reused. The colour of the points correspond to importance sampling coefficient $\varqisw$ of the mixture component $j$ at iteration $i$.}
    \label{fig:ABRIS_sampling}
\end{figure}

The resulting sampling criterion tries to construct a holistic view of the current sample set $\sseti$ by including empirical metrics related to importance sampling and variational inference as well as periodically enforcing model calls (see conditions in \cref{alg:abris_sampling_loop}). This results in numerous sampling-free iterations, see \cref{fig:ABRIS_sampling}, yielding cheap parameter updates of \eqref{eq:stochastic_optimizer}. Considering that the desired use case consists of expensive probabilistic models, this leads to a significant speed-up in inference time without knowledge of the inner workings of the model itself. 

Besides requiring new samples, the second purpose of the update strategy consists of selecting which of its samples $\lvs$ are used in equation \eqref{eq:is_elbograd_MC}. In the case of a Gaussian mixture model (GMM) variational distribution, Arenz et al. \cite{Arenz2020} proposed a random sample selection procedure depending on its components and the number of times a sample was already used. However, in literature, it is common to use the samples of a fixed number $m$ of previous iterations at which the model was evaluated \cite{Uchibe2018, Sakaya2017, Arenz2018}, discarding older iterations. Due to its resemblance to moving averages, we refer to such approaches as moving window approaches. If the probabilistic model is not sampled at every iteration $i$, i.e., since the sampling conditions were not met, the sample window consists of samples from non-sequential iterations \cite{Sakaya2017}. We build upon this idea to construct a moving window approach that naturally regulates its batch sizes $N^i=\card{\Xi^i}$ and the number of distinct components in \eqref{eq:is_distribution} for each iteration $i$. Based on the three sampling criteria elaborated previously, we assess the quality of the existing sample set in an inner loop  referred to as the sampling loop. If $\sseti$ is deemed unable to produce a suitable approximation in a sampling loop iteration, the probabilistic model is evaluated on a batch of $N$ new samples $\lvs \sim \varqfi$. Afterward, the sets $\sseti$, $\mseti$, and $\pseti$ are updated in a moving window approach of window size $m\cdot N$. In other words, as long as the number of batches in the adaptive set $\sseti$ is smaller than $m$, the new batches $\Xi^{i,r}$, $\Phi^{i,r}$ and $\varpi$ are added to their respective sets. If $\card{\sseti}> m \cdot N$, the oldest batch, and its evaluation are removed, keeping the number of batches constant. In case the updated sample set is adequate, the sampling loop is exited, and the \texttt{update\_sets} in \cref{alg:is_BBVI} returns the updated $\sseti$, $\mseti$ and $\pseti$ for \eqref{eq:is_elbograd_MC}. Hence, the window moves with sampling loop iterations $r$ across iterations $i$. Consequently, the number of distinct distributions in $\pseti$ is coupled to the sizes of all batches $\ssetj\in \sseti$. 

Since the initial parameters $\varp^0$ tend to produce a poor approximation to the true posterior distribution, the maximum size of the moving window is set to $\tilde{m}=i$ if $m>i$, where $i$ is the iteration number, to avoid a large amount of model calls at the beginning of the optimization, wasting computational resources. For ease of notation, the tilde is mostly dropped, as the total number of optimization iterations tends to be larger than the maximum window size $m$, such that most of the time $\tilde{m}=m$.

\begin{algorithm}
\caption{Update of the sets $\sseti$, $\mseti$ and $\pseti$ for the importance sampling estimate at iteration $i$ for the adaptive batch-sequential regulated importance sampling approach.}\label{alg:abris_sampling_loop}
\begin{algorithmic}
\Require $i$, $m$, $N$, $N_\text{periodic}$, $\varpi$, $\sseti$, $\mseti$, $\pseti$
\State \texttt{periodic\_sampling} = $\text{mod}(i,N_\text{periodic})==0$\\
\State  $\tilde{m} = \min(m, i)$\\
\For{$r \in \{1 ... \tilde{m}\}$}
\If{$\ess(\sseti, \varpi) \leq N$ \Comment{Deteriorated ESS}\\
$\quad \ $ \textbf{or} $\matrixnorm{\boldsymbol{e}_\text{is}(\sseti, \pseti, \varpi)}{\boldsymbol{A}}> \alpha_\text{sc} \matrixnorm{\boldsymbol{e}_\text{ref}(\varpi)}{\boldsymbol{A}}$ \Comment{Insufficient estimate of \eqref{eq:score_function_property}}\\
$\quad \ $ \textbf{or} \texttt{periodic\_sampling}}\Comment{Periodically}

    \State $\Xi^{i,r}=\Big\{\lvs \ \Big| \ \lvs \sim \varqfi\Big\}_{s=1}^{N}$ \Comment{Draw new samples}
    \State $\Phi^{i,r}=\Big\{\ln\unpostfs\Big\}_{\lvs \in \ \Xi^{i,r}}$ \Comment{Evaluate the probabilistic model}\\
    
    \State $\sseti \gets \sseti \cup \Xi^{i,r}$ \Comment{Add new samples}
    \State $\mseti \gets \mseti \cup \Phi^{i,r}$ \Comment{Add new model calls}
    \State $\pseti \gets \pseti \cup \{\varpi\}$ \Comment{Add current variational parameters}\\

    \If{$\card{\sseti}> \tilde{m} \cdot N$} \Comment{Moving window approach}
        \State delete oldest batch from $\sseti$, $\mseti$ and $\pseti$ 
    \EndIf\\
    \State \texttt{periodic\_sampling} = \texttt{false} \Comment{Only once per $i$, not $r$}
\Else
    \State exit loop \Comment{Adequate sample sets}
\EndIf
\EndFor
\end{algorithmic}
\end{algorithm}

In the presented algorithm, called adaptive batch-sequential regulated importance sampling (ABRIS), the model is always called in batches of $N$ samples. This is done to avoid long idle times of the resources (assuming $N \propto n_\text{resource}$). The sampling loop is, therefore, batch sequential and leads to varying wall times per iteration $i$, depending on the number of model evaluation batches. Although this seems undesirable, given that the difference between $\varq{\lv|\varp^{i+1}}$ and $\varqfi$ tends to be small, the sampling loop can oftentimes be immediately exited without model evaluations, leading to cheap posterior updates. Empirically, 
even if sample batches are evaluated, the number of sampling loop iterations commonly remains one, such that this has little effect on the wall time. In demanding iterations where $m$ sampling loop iterations are needed, the algorithm reverts to classical BBVI without IS but with an increased batch size of $N^i = m \cdot N$. Hence, the expectation \eqref{eq:is_elbograd_MC} for this iteration is done using a large number of new samples, leading to an accurate gradient estimate, regulating and stabilizing the optimization process. 

As a last ingredient, we need a procedure to compute the importance sampling weights $\varqisw$ to complete the ABRIS algorithm. Similar to \cite{Sakaya2017, Arenz2018, Arenz2020}, the relative batch sizes of the mixture components are used to compute the mixture coefficients
\begin{equation}
 \varqisw=\frac{\card{\ssetj}}{\card{\sseti}}.
\end{equation}
This approach does not require additional parameters and gives components with more samples more weight. Since by design, the number of model batches in $\card{\sseti}$ can not decrease over iteration $i$, in later iterations, for new samples to outweigh older ones, more sampling iterations of the inner loop are required. This naturally increases the quality demands on the sample set with an increasing number of parameter updates. The moving window approach keeps this effect bounded such that in the most expensive scenario, $\sseti$ consists of $m\cdot N$ samples of $\varqfi$.
The resulting approach, called adaptive batch-sized regulated importance sampling (ABRIS) BBVI, is tailored to black box Bayesian inverse problems with expensive probabilistic models. From a performance aspect, the algorithm is constructed in a batch-sized manner to harness computational resources. Although score function estimators do not require the derivatives of the probabilistic model, they do require the score function of the variational distribution. In contrast to $\unpostf$, these can oftentimes be obtained based on automatic differentiation, particularly in modern probabilistic libraries. Additionally, it should be mentioned that the score function estimators can not be employed in case its parameterization $\varp$ defines the support of $\lv$ \cite{Mohamed2020}. The same conditions apply to all importance-sampling-based estimators \eqref{eq:is_elbograd_MC} as otherwise its weights might not be bounded. Employing an IS scheme introduces an additional requirement on the variational distribution also originating from the computation of the weights. Here, the variational distribution needs to be evaluated at arbitrary samples of its support. Hence, in case the PDF of the variational distribution is based on a transformation $\lvs = T(\varpi, \sams{\xi})$ from a base distribution $\pdf{\xi}$, the inverse transformation is required to evaluate the density. As long as the transformation is invertible, an importance sampling approach can be employed \cite{Sakaya2017}, although eventually requiring an iterative approximation to identify $\sams{\xi}$. 

Other bounds have been employed in VI \cite{Ranganath2016, Arenz2018, Arenz2020}. Importance sampling approaches to reuse model calls have also been used in gradient-based estimators variational inference approaches \cite{Sakaya2017} and notably in policy search methods \cite{Uchibe2018, Tirinzoni2019}.
Uchibe \cite{Uchibe2018} proposed an optimization approach for the mixture coefficients by minimizing the Kullback-Leibler divergence between the optimal importance sampling distribution and the mixture proposal. In pursuit of stability in gradient estimate, Arenz et al. \cite{Arenz2020} used the difference between the desired number of samples and ESS to dynamically adapt the number of samples for each component of a GMM-based variational distribution. To increase expressiveness in the variational distribution, different variational families such as mixtures \cite{Arenz2018, Arenz2020, Bilionis2014}, variational programs \cite{Ranganath2016}, normalizing flows \cite{Rezende2015, Kobyzev2020, Papamakarios2021} and other \cite{Tran2016, Saeedi2017, Naesseth2018} have been employed for variational inference. 

\section{Numerical Examples}
In this section, we investigate the performance of the ABRIS-BBVI approach. First, we look into performance studies related to density match cases to study performance gain compared to plain BBVI. Afterward, the approach is benchmarked on a random field inference problem and compared against other black box approaches. Finally, the presented algorithm is employed on a multiphysics battery problem. All of the algorithms were implemented in the Python-based open-source code QUEENS \cite{queens}.
\subsection{Gaussian Density Matching}
The first study consists of matching Gaussian distributions. This scenario is suitable for studying the effects of sample size $N$ and window size $m$ for the ABRIS approach for various dimensions. The variational distribution is a normal mean field
\begin{align} \label{exeq:variational_distributon_normal}
    \varqf=\nd{}{\lv|\boldsymbol{\mu} = \varp_\mu,\boldsymbol{C}=\text{diag}\left(\exp(2\varp_C)\right)},
\end{align}
and the target distribution, i.e., the probabilistic model is given by
\begin{align}
    \posterior = \nd{}{\lv|\boldsymbol{\mu}^\text{opt}=\boldsymbol{0},\boldsymbol{C}^\text{opt}=0.1\boldsymbol{I}}.
\end{align}
The dimension of $\lv$ is chosen as $\dim{\lv}=2^p$, the number of samples per iteration $N=2^n$ where $p,n \in \{0,1,2,3,5\}$. The window size is $m \in \{10,20,30,40,50\}$. Adam \cite{Kingma2014} is used as a stochastic optimizer, where its initial learning rate is set to $\alpha_\text{init}=\frac{0.1}{\dim{\lv}}$. The number of periodic evaluation iterations equals $N_{\text{periodic}}=50$. For $m=0$, using $N_{\text{periodic}}=1$, the original form of BBVI, i.e.~without IS, is recovered. For all the examples, the ELBO gradient was preconditioned using the inverse of the Fisher information matrix (FIM) of the variational distribution:
\begin{align}
    \boldsymbol{F}(\varpi) =   \begin{bmatrix}
  \text{diag}(\exp(-2\varp_C^i)) & \boldsymbol{0} \\
  \boldsymbol{0} & 2 \boldsymbol{I} 
  \end{bmatrix}
\end{align}
To ensure numerical invertibility of the FIM \cite{Tang2019} we add an exponential damping $\boldsymbol{\tilde{F}}(\varpi)=\boldsymbol{F}(\varpi)+\eta^i \boldsymbol{I}$ where
\begin{equation}
    \eta^i= \begin{cases}
        \tilde{\eta} & \text{if }i< i^{b} \\
        \max(\tilde{\eta} \cdot \exp(-\frac{i-i^{b}}{i^{b}}), \eta_\text{bound}) & \text{else}
    \end{cases}        
\end{equation}
For these numerical examples $\tilde{\eta}=10^{-2}$, $i^{b}=50$ and $\eta_\text{bound}=10^{-6}$. The ELBO gradients are clipped by the $L_2$-norm with a threshold of $10^6$. In all examples, the baseline parameter is set to
\begin{equation}
   a=\frac{\sum_{c=1}^{n_{p}}\Cov{\varqfi}{\score{\lambda_c}{\varqfi},\boldsymbol{g}^\text{sc}_c(\lv,\varpi)}}{\sum_{c=1}^{n_{p}}\Var{\varqfi}{\score{\lambda_c}{\varqfi}}}, 
\end{equation}
where the expectations are approximated with Monte Carlo using $\mseti$, $\sseti$ and $\pseti$. Although this is not the theoretical parameter as described in \cite{Ranganath2014}, empirically, it provided stable results for the investigated examples. The parameters are uniformly randomly initialized in $[-0.1,0.1]$ for $\varp_\mu$ and $[\ln(0.4), \ln(0.2)]$ for $\varp_C$ respecitvely.

By design, the distribution can be matched exactly. This is done to focus on performance properties for a fixed accuracy. The optimization is stopped once the $L_2$-norm of the error between current and optimal variational parameters relative to the norm of the reference value falls below $10^{-3}$, $\niter\geq 150000$ or the number of model calls exceeds $10^6$. For each $m$, $N$, and $p$, the experiments are repeated ten times with different random initializations of the variational parameters and random seeds. This experiment aims to evaluate the performance, specifically the accuracy per probabilistic model call.
\begin{figure}[h!]
    \centering
    \input{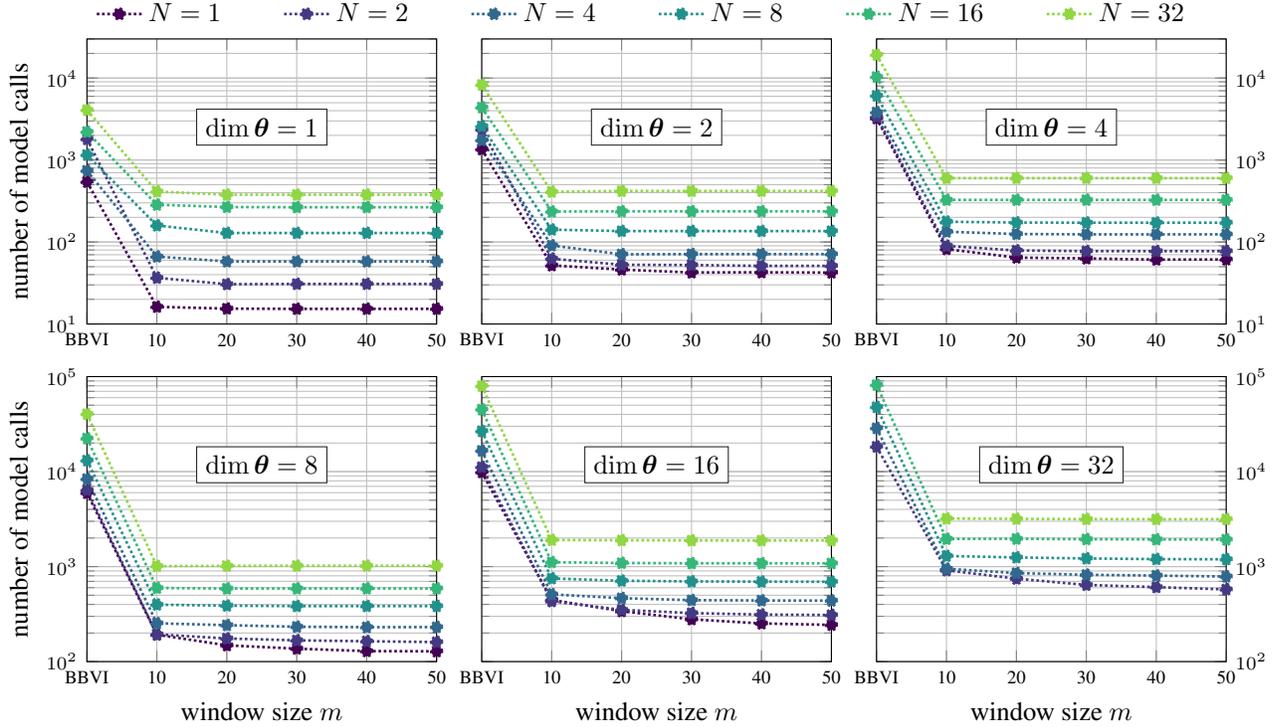}
    \caption{Gaussian match cases for $\dim{\lv}\in \{1,2,4,8,16,32\}$. The number of model calls is plotted against window size $m$ for various samples per iteration $N$. The number of simulations is the average of 10 runs with different random seeds. BBVI indicates that no importance sampling was used.}
    \label{fig:nmc}
\end{figure}

For $\dim{\lv}\leq 8$, the ABRIS cases consistently outperform their original BBVI counterpart for a fixed $N$, see \cref{fig:nmc}. Up to dimension 8, the algorithm behaves similarly for all the cases. Here, the ABRIS approach can drastically reduce the number of model calls under the same conditions. Also, it can be seen that the window size has little effect on the required model calls, making it a simple parameter to set up. For in every case, for $N\geq 4$, all the plots \cref{fig:nmc} display an approximate constant number of model evaluations solely depending on $N$. For $\dim{\lv}=16$, lower middle plot in \cref{fig:nmc}, if $N=1$ and $m=10$, most runs were not able to converge within 150000 iterations. However, this was fixed by increasing the window size ($\geq 20$). Here, the adaptivity stabilized the inference algorithm as it allows sampling more from the current distribution if needed. Even though a single sample per iteration converged for this density match case for the classical BBVI, it is generally challenging to tune the stochastic optimizer to convergence. For the cases $\dim{\lv}=32$ and $N=1$, most runs were not able to converge with the prescribed setting. Therefore, using batch sizes larger than a single sample is recommended. In the $16$-dimensional setting, moving to large samples per iteration made all examples converge. A similar picture is seen for the $32$-dimensional case in the lower right plot of \cref{fig:nmc}. Notably, it can be observed that larger window sizes can facilitate the optimizers' convergence. Due to the limit on model calls, the cases $\dim{\lv}=32$ with plain BBVI and $N=1$ are not able to converge within $10^6$ model calls and are therefore omitted in \cref{fig:nmc}.

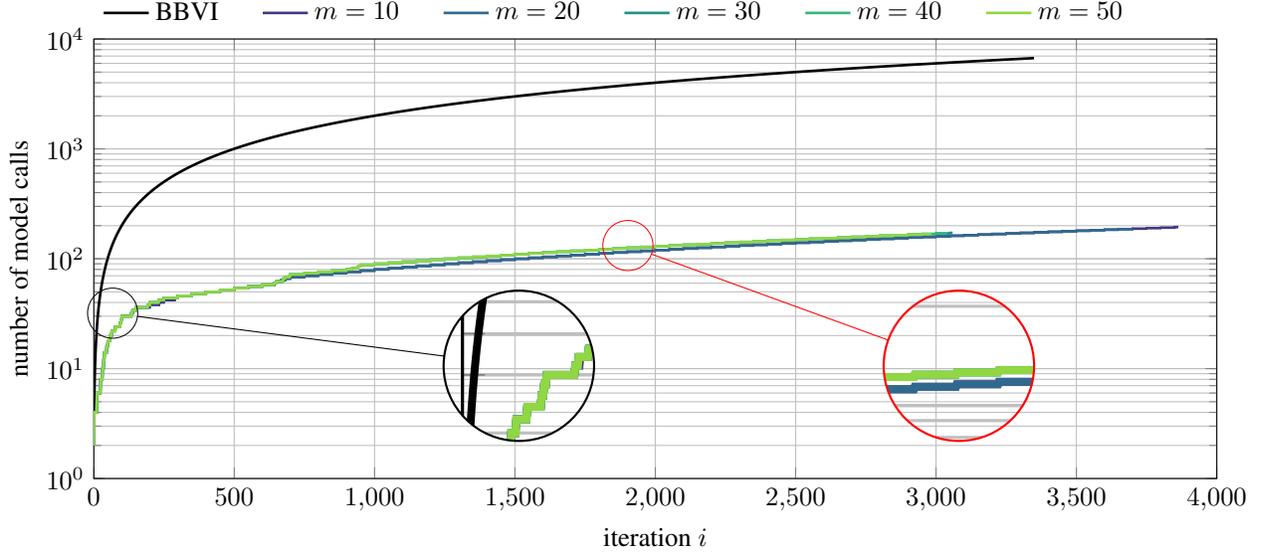
\begin{figure}[h]
    \centering
    \usetikzlibrary[spy]
\pgfplotsset{compat=newest}
\begin{tikzpicture}[spy using outlines={circle,size=2cm, connect spies}]
    \definecolor{m0}{RGB}{68, 1, 84}
    \definecolor{m10}{RGB}{68, 57, 131}
    \definecolor{m20}{RGB}{49, 104, 142}
    \definecolor{m30}{RGB}{33, 145, 140}
    \definecolor{m40}{RGB}{53, 183, 121}
    \definecolor{m50}{RGB}{144, 215, 67}
  \begin{groupplot}[
    group style={
      group size=1 by 1, 
      horizontal sep=1cm, 
    },
    colormap/viridis,
    width=\textwidth,
    height=0.45\textwidth
  ]

\nextgroupplot[
    xlabel={iteration $i$},
    ylabel={number of model calls},
    ymode = log,
    xmin=0, xmax=4000,
    xtick={0,500,1000,1500,2000,2500,3000,3500,4000},
    ymin=1, ymax=10000,
    ytick={1,10,100,1000,10000},
    xmajorgrids,
    ymajorgrids,
    yminorgrids,
    legend style={at={(0,1.01)},legend columns=-1, anchor=south west,draw=none, /tikz/every even column/.append style={column sep=0.5cm}},
  ]
 \addplot[color=black, line width=1] table [x=iteration, y=n_sims, col sep=comma] {figures/nmc/n_sims_8_2_0.csv}; \addlegendentry{BBVI}
 \addplot[color=m10, line width=1] table [x=iteration, y=n_sims, col sep=comma] {figures/nmc/n_sims_8_2_10.csv}; \addlegendentry{$m=10$}
 \addplot[color=m20, line width=1] table [x=iteration, y=n_sims, col sep=comma] {figures/nmc/n_sims_8_2_20.csv}; \addlegendentry{$m=20$}
 \addplot[color=m30, line width=1] table [x=iteration, y=n_sims, col sep=comma] {figures/nmc/n_sims_8_2_30.csv}; \addlegendentry{$m=30$}
 \addplot[color=m40, line width=1] table [x=iteration, y=n_sims, col sep=comma] {figures/nmc/n_sims_8_2_40.csv}; \addlegendentry{$m=40$}
 \addplot[color=m50, line width=1] table [x=iteration, y=n_sims, col sep=comma] {figures/nmc/n_sims_8_2_50.csv}; \addlegendentry{$m=50$}
 \spy[black,magnification=3]  on (0.25,2.2) in node at (5.65,1.5);
 \spy[red,magnification=3]  on (7.1,3.1) in node at (11.5,1.5);
\end{groupplot}  
\end{tikzpicture}
    \caption{Number of simulations plotted over the optimiser iterations for $\dim{\lv}=8$ with $N=2$ for various window sizes $m$. In the black zoom-in, the adaptive sampling strategy is visible. The red zoom-in shows the dominating periodic condition towards the end of the optimization.}
    \label{fig:nmc_n_sims_8d}
\end{figure}

In \cref{fig:nmc_n_sims_8d}, the reduction in model calls becomes apparent. For the base algorithm, the number of simulations is $N \cdot N_\text{iter}$, whereas the ABRIS cases require approximately two orders of magnitude less model calls.
For the eight-dimensional match case shown in \cref{fig:nmc_n_sims_8d}, the sampling loop consists of at most two iterations, i.e., the number of model calls in a single iteration $i$ is increased by $2 \cdot N=4$. The black zoom-in in \cref{fig:nmc_n_sims_8d} highlights the adaptive nature of the ABRIS algorithm, permitting an increased number of model calls in certain iterations. Furthermore, it can be seen that the periodic condition is dominating  towards the end of the optimization (red zoom-in in \cref{fig:nmc_n_sims_8d}). This is an essential part of the ABRIS approach as it enforces exploration of the parameter space, even in the vicinity of a local optimum. Here, the IS sample sets tend to be a good approximation for the expectation, such that no new samples are required, leading to virtually free parameter updates.

\subsection{Generalized Poisson Calibration Problem}
\begin{figure}[h!]
    \centering
    \input{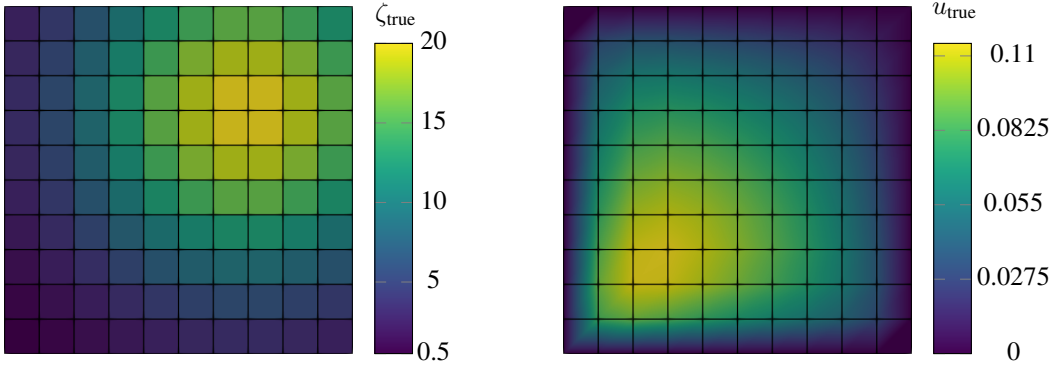}
    \caption{Ground truth $\zeta_\text{true}$ (left) and resulting model output (right).}
    \label{fig:gle_ground_truth}
\end{figure}

In this numerical example, we investigate the calibration of a constitutive field of a generalized Poisson problem, commonly needed in fields such as biomechanical~\cite{Hervas-Raluy2023}, electrochemical~\cite{Schmidt2023}, or thermal systems~\cite{Proell2024}. In this numerical study, the forward problem is defined by:
\begin{align}\label{eq:laplace_problem}
    \nabla \cdot (-\zeta \nabla u)&=10 \text{ in } \Omega, \\
    u &= 0 \text{ on } \Gamma_D.
\end{align}
The three-dimensional domain $\Omega$ consists of a box with sides $[-0.05,0.05] \times [-0.5,0.5] \times [-0.5,0.5]$ and Dirichlet condition is prescribed on the boundary $\Gamma_D=\{|x_2|=0.5 \text{ or }  |x_3|=0.5\}$. The system is solved numerically using a finite element approach with $1 \times 10 \times 10$ tri-linear elements. The forward model outputs $\DM(\lv)=\boldsymbol{u}^h(x_1 = -0.05)$ for a given $\lv$ value is the 121-dimensional nodal solution vector. The inverse problem consists of the calibration of the spatially variable coefficient $\zeta$, based on noisy observations $\yobs$ of the nodal solution vector $\boldsymbol{u}^h_\text{true}$. In each element $e$, the true value $\zeta_\text{true}$, left plot in \cref{fig:gle_ground_truth}, is given by
\begin{align}
    \zeta_\text{true}(\boldsymbol{x}_\text{e}) = 20 \exp\left(-4\norm{\boldsymbol{x}_\text{e}-(0,0.2,0.2)^T}^2\right).
\end{align}
where $\boldsymbol{x}_\text{e}$ denotes the center of the quadrilateral element $e$. The corresponding observations $\yobs$ are generated by 
\begin{align}
    \yobs = \boldsymbol{u}^h_\text{true} + 10^{-3} \overline{u}^h_\text{true} \sample{\boldsymbol{\xi}}{s},
\end{align}
where $\overline{u}^h_\text{true}$ is the mean value of $\boldsymbol{u}^h_\text{true}$ (right plot in \cref{fig:gle_ground_truth}) and $\sample{\boldsymbol{\xi}}{s}$ a sample of a standard Gaussian random variable.

For the inference process, the random field $\zeta$ is modeled based on a truncated Kosambi-Karhunen-Lo\`{e}ve expansion \cite{Kosambi1943,Stefanou2007} with $n_\text{KKL}$ terms
\begin{equation}\label{eq:KKL}
    \ln\zeta(\lv,\boldsymbol{x}_\text{e}) = \sum_{j=1}^{n_\text{KKL}} \lvj \phi_j(\boldsymbol{x}_\text{e}).
\end{equation}
The basis functions $\phi_j(\boldsymbol{x})$ are constructed based on the eigenfunctions of a Gaussian random field using a squared exponential (SE) kernel  with a length scale of $0.3$. The eigenfunctions are discretized numerically using the eigenvalue decomposition of the SE kernel Gram matrix evaluated at the element centers. The propagation through an exponential function in \eqref{eq:KKL} ensures positivity of $\zeta$. For all the examples, we set the dimension of the latent variable to $\dim{\lv}=n_\text{KKL}=20$.

The probabilistic model consists of a standard Gaussian prior on $\lv$ as well as a Gaussian likelihood:
\begin{equation}
    \unpostf =\nd{}{\yobs|\DM(\lv), \boldsymbol{C}} \nd{}{\lv|\boldsymbol{0}, \boldsymbol{I}},
\end{equation}
using the diagonal covariance matrix
\begin{align}
    C_{jj} = \left(10^{-2}|{y_{\text{obs},j}}|+10^{-2}\right)^2.
\end{align}
This choice alters the variance of each component of the nodal solution vector in order to accommodate the large spread in values in $\yobs$ whilst avoiding too small values using an offset by $10^{-2}$. Note that this choice is distinct from the noise added to $\boldsymbol{u}^h_\text{true}$. 

This study compares the ABRIS-BBVI approach to Markov Chain Monte Carlo (MCMC) and Sequential Monte Carlo methods (SMC). For the MCMC cases, a Metropolis-Hastings scheme is chosen with a Gaussian proposal distribution. The covariance of the proposal distribution is scaled at every $i_\text{tune}$th step to reach an acceptance rate within $[0.2, 0.5]$. A burn-in of half the amount of desired samples is used. The algorithm was implemented in the open-source code QUEENS. In the sequential Monte Carlo runs, the number of particles is defined by $N_\text{particles}$ and $i_\text{rejuvenation}$ rejuvenation steps are used within an SMC step. A waste-free approach using residual resampling with a threshold of 0.5 and an adaptive tempering scheme \cite{Dau2022} is employed. The SMC implementation is based on the python package \textit{particles} \cite{Chopin2024}, where the likelihood model is implemented in QUEENS. This allows the exploitation of parallel infrastructure using batches of simulations during the rejuvenation steps. Since the adaptive tempering makes it impossible to set a specific number of desired model calls $N_\text{m}$, we select the initial rejuvenation steps to $i_\text{rejuvenation}=10$ and particle number to $N_\text{particles}=100$. In case the SMC run failed, both these parameters were doubled, and the SMC run restarted. This highlights the difficulties of setting these parameters for small computational budgets. The model calls of the failed simulations are not included in the total model calls below. For the ABRIS cases, the variational distribution is chosen as mean-field Gaussian. The window size is set to $m=10$, and the batch size is set to $N=8$. The learning rate of the stochastic optimizer is reduced by a factor of 0.9 every 1000th iteration to start with a larger initial value. The number of periodic evaluations is $N_{\text{periodic}}=100$. For every method, each run was repeated 5 times. For all the cases except MCMC, two simulations were done in parallel. 

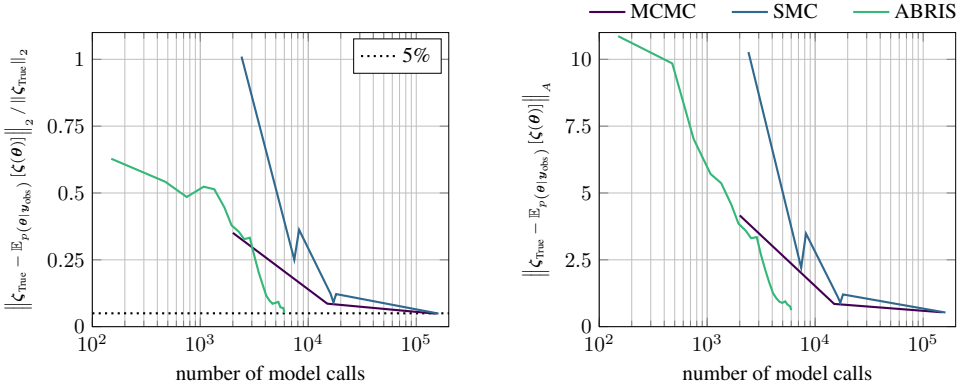
\begin{figure}[H]
    \centering
    \pgfplotsset{
    discard if not/.style 2 args={
        x filter/.code={
            \edef\tempa{\thisrow{#1}}
            \edef\tempb{#2}
            \ifx\tempa\tempb
            \else
                \def\pgfmathresult{inf}
            \fi
        }
    }
}
\pgfplotsset{compat=newest}
\begin{tikzpicture}[scale=0.8]
    \definecolor{MCMC}{RGB}{68, 1, 84}
    \definecolor{SMC}{RGB}{49, 104, 142}
    \definecolor{ABRIS}{RGB}{53, 183, 121}
  \begin{groupplot}[
    group style={
      group size=2 by 1, 
      horizontal sep=2.5cm, 
    },
    colormap/viridis,
    width=\textwidth/2.2,
  ]
   \nextgroupplot[
    xlabel={number of model calls},
    xmode = log,
    xmin=100, xmax=200000,
    xtick={100,1000,10000,100000},
    ylabel={$\norm{\boldsymbol{\zeta}_\text{True}-\Ex{\posterior}{\boldsymbol{\zeta}(\lv)}}_2 / \norm{\boldsymbol{\zeta}_\text{True}}_2$},
    ylabel style = {font=\scriptsize},
    ymin=0.0,
    ymax=1.1,
    ytick={0,0.25,0.5,0.75,1},
    xmajorgrids,
    xminorgrids,
    ymajorgrids,
  ]
  
  \addplot[dotted, color=black,mark size = 1pt, line width=1] coordinates {(100,0.05)(1000000,0.05)};\addlegendentry{5\%}; 
  \addplot[color=MCMC, mark size = 2pt, line width=1] table [x=n_sims, y=mean_error, col sep=comma] {figures/gle/laplace_averaged_MCMC.csv};
  \addplot[color=SMC, mark size = 2pt, line width=1] table [x=n_sims, y=mean_error, col sep=comma] {figures/gle/laplace_averaged_SMC.csv};
  \addplot[color=ABRIS, mark size = 2pt, line width=1] table [x=n_sims, y=mean_error, col sep=comma] {figures/gle/laplace_averaged_ABRIS_2.csv};
\nextgroupplot[
    xlabel={number of model calls},
    xmode = log,
    xmin=100, xmax=200000,
    xtick={100,1000,10000,100000},
    ylabel={$\norm{\boldsymbol{\zeta}_\text{True}-\Ex{\posterior}{\boldsymbol{\zeta}(\lv)}}_A$},
    ylabel style = {font=\scriptsize},
    ymin=0.0,
    ymax=11,
    ytick={0,2.5,5,7.5,10},
    xmajorgrids,
    xminorgrids,
    ymajorgrids,
    legend style={at={(-0.05,1.01)},legend columns=-1, anchor=south west,draw=none, /tikz/every even column/.append style={column sep=0.5cm}},
  ]

\addplot[color=MCMC, mark size = 2pt, line width=1] table [x=n_sims, y=lv_error, col sep=comma] {figures/gle/laplace_averaged_MCMC.csv}; \addlegendentry{MCMC}; 
\addplot[color=SMC, mark size = 2pt, line width=1] table [x=n_sims, y=lv_error, col sep=comma] {figures/gle/laplace_averaged_SMC.csv}; \addlegendentry{SMC}
\addplot[color=ABRIS, mark size = 2pt, line width=1] table [x=n_sims, y=lv_error, col sep=comma] {figures/gle/laplace_averaged_ABRIS_2.csv}; \addlegendentry{ABRIS}; 

\end{groupplot}  
\end{tikzpicture}
    \caption{Comparison between posterior mean values and the ground truth $\boldsymbol{\zeta}_\text{true}$ in relative $L_2$-error and in $\boldsymbol{A}$-norm, weighting each elementwise constant coefficient with $A_{jj}=\zeta_\text{true, j}^2$, for the generalized Poisson problem.
    }
    \label{fig:gle_mean_values}
\end{figure}

\begin{figure}[H]
    \centering
    \input{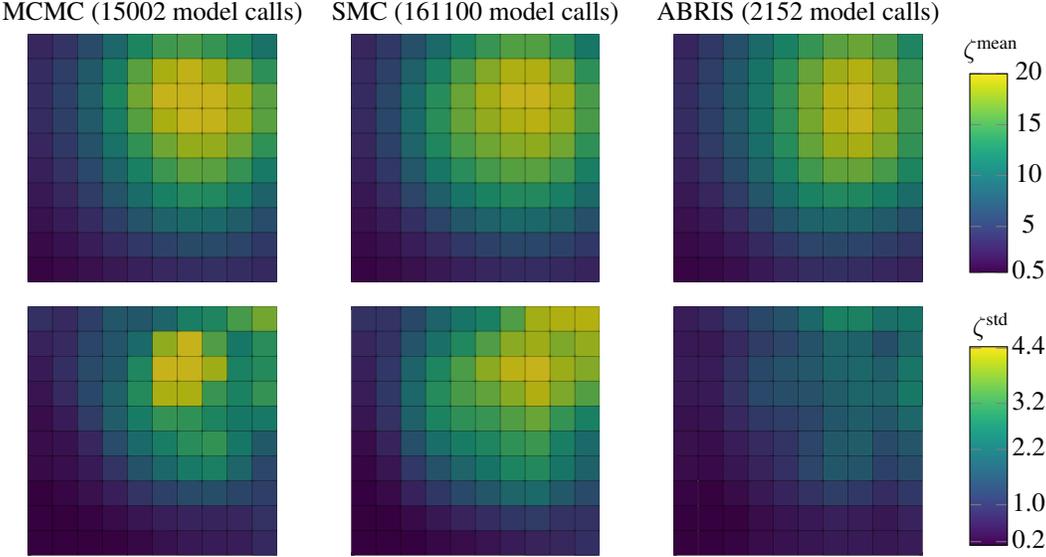}
    \caption{Comparison of mean values $\zeta^\text{mean}$ (top row) and standard deviations $\zeta^\text{std}$ (bottom row) for the different algorithms. In each case, the run where the relative error to the ground truth is approximately five percent, with the lowest amount of forward calls is shown.}
    \label{fig:gle_field}
\end{figure}

Averaged over the 5 runs,  ABRIS-BBVI is able to outperform MCMC and SMC in terms of model calls for a fixed accuracy, once the number of model calls exceeds 2000. This can be seen for two different error norms in \cref{fig:gle_mean_values}. For a relative $L_2$ error of approximately five percent in the posterior mean (averaged of all five runs per inference scheme), the ABRIS cases require 6000 simulation calls, outperforming MCMC with 150002 and SMC with 161000, respectively. Notably, this reduction is achieved without changing the probabilistic model itself, instead operating directly on the expensive forward model. For the MCMC cases, using multiple chains led to convergence issues, such that a single chain was used. This increases the wall time spent in model calls as parallelism could not be exploited. Due to resource constraints, eight simulations were done in parallel for the SMC and ABRIS cases. Therefore, the ABRIS approach outperforms SMC by a factor of $\approx 26$ and MCMC by a factor of $\approx 200$ in terms of model calls averaged over the five random seeds in \cref{fig:gle_mean_values}. The posterior mean and standard deviations for the run with the least amount of model calls to reach a relative $L_2$-error to the ground truth of approximately five percent is show in \cref{fig:gle_field}.

\subsection{Bayesian Calibration of a Physics-Based Solid-State-Battery Model}
The ABRIS approach is primarily suited for non-differentiable, expensive probabilistic models constrained by a limited computational budget. This numerical example focuses on Bayesian calibration with a solid-state battery (SSB) simulation model to demonstrate the algorithm in such a setting. Solid-state lithium-ion batteries are a novel concept for energy storage, expected to circumvent the physical limitations of conventional liquid electrolyte lithium-ion cells \cite{Janek2016}. 

\begin{figure}[h]
    \centering
    \begin{tikzpicture}
\def\radius{2}
\def\costheta{0.96593}
\def\sintheta{0.25882}
\def\refwidth{14}
\def\refthickness{3}
\def\ccthickness{1.2}
\def\anthickness{2.2}
\draw[thick] (0,0) rectangle (\refwidth,\refthickness);
\draw[brown, thick] (0,0) -- (0,\refthickness);
\draw[cyan, thick] (\ccthickness,0) -- (\ccthickness,\refthickness);
\draw[green, thick] (\ccthickness+\anthickness,0) -- (\ccthickness+\anthickness,\refthickness);
\draw[blue, thick] (\refwidth-\ccthickness,0) -- (\refwidth-\ccthickness,\refthickness-\sintheta*\radius);
\draw[red, thick] (\refwidth-\ccthickness,\refthickness-\sintheta*\radius) -- (\refwidth-\ccthickness,\refthickness);
\draw[olive, thick] (\refwidth,0) -- (\refwidth,\refthickness);

\draw[<->] (\ccthickness+\anthickness, \refthickness+0.15) -- (\refwidth-\ccthickness-\refthickness*\costheta*\radius-\radius, \refthickness+0.15) node[midway, above]{$d_\text{ses}$} ;

\draw[violet, thick] (\refwidth-\ccthickness-\costheta*\radius,\refthickness) ++(195:\radius) arc (195:345:\radius);
\draw[violet, thick] (\refwidth-\ccthickness-3*\costheta*\radius,\refthickness) ++(180:\radius) arc (180:345:\radius);

\def\domainsymbolheight{1.5}
\node at (0.5*\ccthickness, \domainsymbolheight) {$\Omega_\text{cc,a}$};
\node at (\ccthickness+0.5*\anthickness,\domainsymbolheight) {$\Omega_\text{a}$};
\node at (4.4, 0.5*\domainsymbolheight) {$\Omega_\text{el}$};
\node at (0.5*\refwidth+2.2*\radius,0.8+\domainsymbolheight) {$\Omega_\text{c}$};
\node at (\refwidth-0.5*\ccthickness, \domainsymbolheight) {$\Omega_\text{cc,c}$};

\def\boundarysymboloffset{0.3}
\node at (0,           -\boundarysymboloffset) {\color{brown}$\Gamma_\text{cc,a-o}$};
\node at (\ccthickness+\anthickness,         -\boundarysymboloffset) {\color{green}$\Gamma_\text{a-el}$};
\node at (10,         0.5*\domainsymbolheight) {\color{violet}$\Gamma_\text{c-el}$};
\node at (\refwidth-\ccthickness, -\boundarysymboloffset) {\color{blue}$\Gamma_\text{cc-el}$};
\node at (\refwidth,   -\boundarysymboloffset) {\color{olive}$\Gamma_\text{cc,c-o}$};
\node at (\ccthickness,           3+\boundarysymboloffset) {\color{cyan}$\Gamma_\text{cc-a}$};
\node at (\refwidth-\ccthickness,           3+\boundarysymboloffset) {\color{red}$\Gamma_\text{cc-c}$};
\node at (7,           3+\boundarysymboloffset) {$\Gamma_\text{cut}$};
\end{tikzpicture}
    \caption{Two-dimensional view of the SSB geometry (wedge) for two active material particles $n_c=2$ and separator thickness $d_\text{ses}$. The domain consists of an anode $\Omega_\text{a}$, electrolyte $\Omega_\text{el}$, a composite cathode with cathode active material particles $\Omega_\text{c}$ sandwiched between two current collectors $\Omega_\text{cc,a}$ and $\Omega_\text{cc,c}$. The electrode domain is defined as $\Omega_\text{ed}=\Omega_\text{a}\cup\Omega_\text{c}$ and the interface to the electrolyte phase is defined by $\Gamma_\text{ed-el} =\Gamma_\text{a-el} \cup \Gamma_\text{c-el}$.}
    \label{fig:battery_domain}
\end{figure}
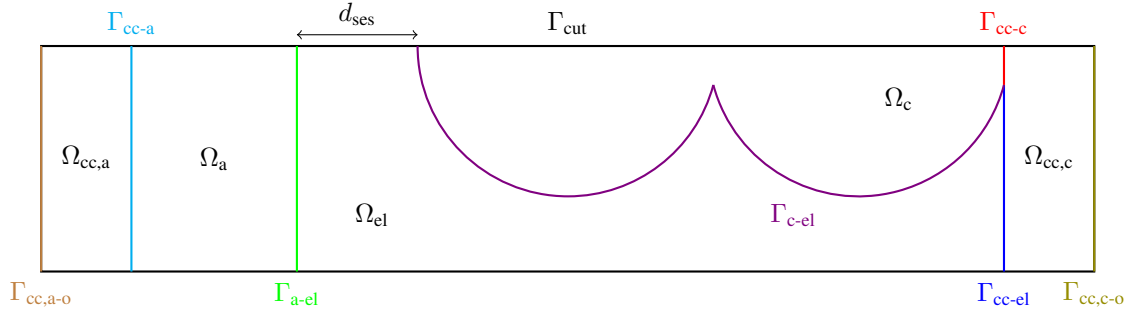

Since they currently exist mainly on a laboratory scale, modeling and calibrating model parameters are crucial to accelerating battery research. For physically sound parameter inference, the probabilistic model includes a physical, PDE-based forward model $\DM(\lv)$ to simulate the electrochemical behavior during dis/charge. The multiphysics continuum-based model couples mechanical with electrochemical behavior. Since the physical model requires fundamental knowledge in various physical fields, we only present the main PDEs related to this numerical example. For a comprehensive derivation of the equations based on adequate modeling assumptions, the reader is referred to \cite{Schmidt2023}. A schematic sketch of an SSB battery cell geometry is depicted in \cref{fig:battery_domain}. Mechanics are modeled using the momentum conservation
\begin{align}
    \grad{\cdot (\boldsymbol{F}\cdot\boldsymbol{S})}{\boldsymbol{X}}+\boldsymbol{b}_0=\rho_0 \acc \text{  in } \Omega_0 \times T
\end{align}
in material configuration $\boldsymbol{X}\in \Omega_0$, where $\boldsymbol{F}$ is the deformation gradient, $\boldsymbol{S}$ the second Piola–Kirchhoff stress tensor, $\boldsymbol{b}_0$ the body force vector, $\rho_0$ the density and $\acc$ the acceleration due to body deformation. In every subdomain, a Neo-Hookean constitutive law is used, parameterized with material-dependent settings. The time domain is the union of charge and discharge time $T=[0, t_\text{charge}]\cup(t_\text{charge},t_\text{charge}+t_\text{discharge}]$. 

Based on the electric potential $\Phi$, the electronic conductivity of the active material $\sigma$,  and the ionic conductivity of the electrolyte $\kappa$, the electrochemical behavior is expressed through the charge conservation 
\begin{align}
    \nabla \cdot (-\sigma \nabla\Phi)&=0 \text{ in } \Omega\setminus \Omega_\text{el}\times T\\
    \nabla \cdot (-\kappa \nabla\Phi)&=0 \text{ in } \Omega_\text{el} \times T,
\end{align}
as well as the mass conservation
\begin{align}
    \nabla \cdot (-D \nabla c) +\pfrac{c}{t}\Big{|}_{\boldsymbol{X}} + c \grad{\cdot \vel}{} &= 0 \text{ in } \Omega_\text{ed} \times T\\
    \pfrac{c}{t}\Big{|}_{\boldsymbol{X}} + c\grad{\cdot \vel}{}&= 0 \text{ in } \Omega_\text{el} \times T.
\end{align}
The concentration of lithium-ions in the electrolyte and lithium in the active material is denoted by $c$. For the latter, the lithium diffusion coefficient is represented by $D$. In the active material, a lithiation-dependent growth rate is applied, resulting in a volumetric coupling between mass conservation and mechanics within that subdomain. The dis/charging of the cell with a C-rate $c_r$ is obtained by prescribing the current density $\boldsymbol{i}$ in normal direction $\boldsymbol{n}$ to the boundary
\begin{align}
    \boldsymbol{i}\cdot\boldsymbol{n}= \begin{cases}
         n_\text{c} \hat{i} c_r &\text{ on } \Gamma_\text{cc,c-o}\times[0, t_\text{charge}]\\
        -n_\text{c} \hat{i} c_r &\text{ on } \Gamma_\text{cc,c-o}\times(t_\text{charge},t_\text{charge}+ t_\text{discharge}]\\
    \end{cases}
\end{align}
where $n_\text{c} \hat{i}$ is the current density needed to cycle the cell with $c_r=1$ for $n_\text{c}$ number of active material particles. The C-rate $c_r$ of 1 defines the current density that needs to be applied to transfer the whole capacity within one hour of dis/charging.  The remaining initial and boundary conditions can be found in the appendix ~\ref{sec:appendix_SSB}. The interaction of the electrochemical fields at the two-dimensional interfaces $\Gamma_\star$, where $\star$ is either an anode-electrolyte $\Gamma_\text{a-el}$ or a cathode active material-electrolyte interface $\Gamma_\text{c-el}$. They are coupled based on a Butler--Volmer approach:
\begin{align}
    i^\star &=i_0^\star \left[\exp\left(\frac{\alpha_a F \eta }{R\vartheta}\right)-\exp\left(-\frac{(1-\alpha_a) F \eta }{R\vartheta}\right)\right] \text{ on } \Gamma_\star \times T\\
    \eta &= \Phi_\text{ed} - \Phi_\text{el} - \Phi_0 \text{ on } \Gamma_\star \times T,
\end{align}
where $i$ is the interface current density, $i_0$ the exchange current density, $F$ the Faraday constant, $R$ the gas constant, $\vartheta$ the temperature, $\alpha_a$ the anodic symmetry coefficient and $\eta$ the overvoltage at the interface $\Gamma_\star$. It models the de/intercalation reaction where $\Phi_\text{ed}$ and $\Phi_\text{el}$ represent the electric potential in their respective domains and $\Phi_0$ the half-cell open circuit voltage.

The set of PDEs, interface, and boundary conditions are discretized using a finite element approach in space and an adaptive one-step-theta scheme to discretize the mass conservation and a Generalized-$\alpha$ scheme employing the same time step size to discretize the momentum conservation in time. The resulting nonlinear system of equations is solved monolithically based on a Newton--Raphson method implemented in the open-source multiphysics code 4C \cite{4C}. More information on the physical modeling and its discretization can be found in the original publication \cite{Schmidt2023}.

As can be seen, solving this system of equations requires in-depth knowledge in various fields, namely physical modeling, coupled systems, numerical discretizations, and high-performance computing. Constructing such a model is a highly complex task resulting from years of experience.

In this section, a forward model call refers to the simulation of an SSB for a charge and discharge cycle, generating three quantities of interest, the cell voltage during the discharge $\DM^\Phi(\lv)=\Phi^\text{cell}$, the state-of-charge $\DM^\text{SOC}(\lv)=\text{SOC}$ and the difference between amount-of-substance of lithium in the active material $\DM^\text{AOS}(\lv)=\Delta\text{AOS}$ during discharge, linearly interpolated at the observed time points. The simulation is stopped if the discharge time exceeds the observed discharge time. In case physical time in the simulation is shorter than the observed one, the model output quantities are kept constant to mimic physical behavior.
 
\begin{figure}[h]
    \centering
    \input{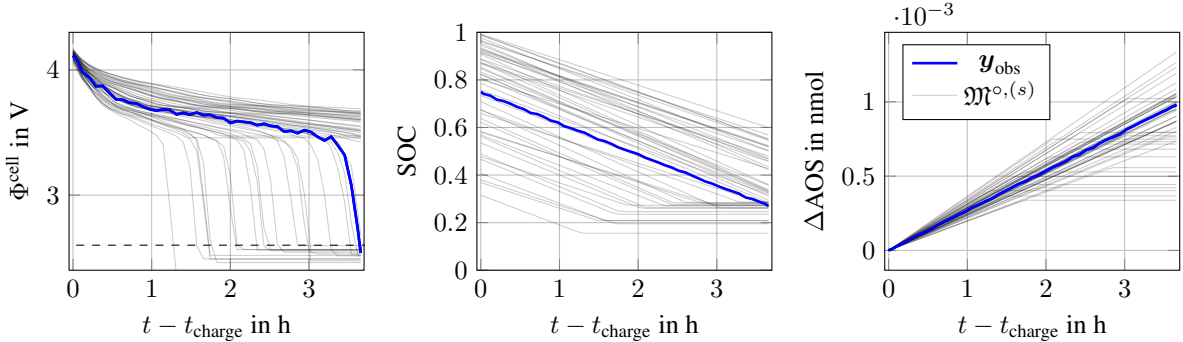}
    \caption{Model outputs and observations over time. On the left is the cell voltage, in the middle the state-of-charge, and on the right the difference in amount-of-substance of lithium over time. The blue line indicates the observational data, whereas the grey lines are model outputs $\DM^{\circ, (s)}$, with $\circ \in \{\Phi, \text{SOC}, \text {AOS}\}$ for different input samples $\lvs$.}
    \label{fig:battery_observations}
\end{figure}
Based on the forward model, the goal of this numerical example is to calibrate four physical parameters (see \cref{tab:SSB_parameters}) from observation data $\yobs$ at the discrete time steps. The observational data $\yobs$ was generated by using the ground truth $\lv^\text{true}$ and corrupting the model output with uniform noise centered around the model output. For the cell voltage observations, the noise range was estimated from the discharge experiments in \cite{Koerver2018}. For the other two quantities of interest, the noise value was set to one percent of the maximum value of the respective quantity. The observational data and possible model outputs can be found in \cref{fig:battery_observations}.

For the Bayesian inverse analysis, we assumed
\begin{align}
    \yobs^{\circ} = \DM^\circ(\lv) + \varepsilon^\circ, 
\end{align}
where the noise $\varepsilon^\circ$, for $\circ \in \{\Phi, \text{SOC}, \text {AOS}\}$, is assumed to be Gaussian. The resulting likelihood yields
\begin{align}
    \like=\nd{}{\yobs^\Phi|\DM^\Phi(\lv), \boldsymbol{C}^\Phi}\nd{}{\yobs^\text{SOC}|\DM^\text{SOC}(\lv), \boldsymbol{C}^\text{SOC}}\nd{}{\yobs^\text{AOS}|\DM^\text{AOS}(\lv), \boldsymbol{C}^\text{AOS}},
\end{align}
where for each field an isotropic covariance matrix $\boldsymbol{C}^\circ = (0.025 \max \yobs^\circ)^2 \boldsymbol{I}$ is used.
\begin{table}[h!]
    \renewcommand{\arraystretch}{1.3}
    \centering
    \begin{tabular}{c|c|c|c}
         parameter $\theta^\text{phy}_j$ & symbol & unit  & sample space $\mathbb{X}_j$\\
         \hline
         diffusion constant of active material & $D$ & $\frac{\text{m}^2}{\text{s}}$  & $[10^{-14}, 10^{-13}]$ \\
         charge C-rate & $c_r$ & $-$  &$[0.1,0.15]$ \\
         separator thickness & $d_\text{ses}$ & $\mu $m  & 
         $\{370,390,410,430\}$ \\
         number of cathode particles & $n_\text{c}$ & $-$   & $\{8,9,10,11\}$\\
    \end{tabular}
    \caption{Parameters to infer for the SSB system and their respective sample space.}
    \label{tab:SSB_parameters}
\end{table}

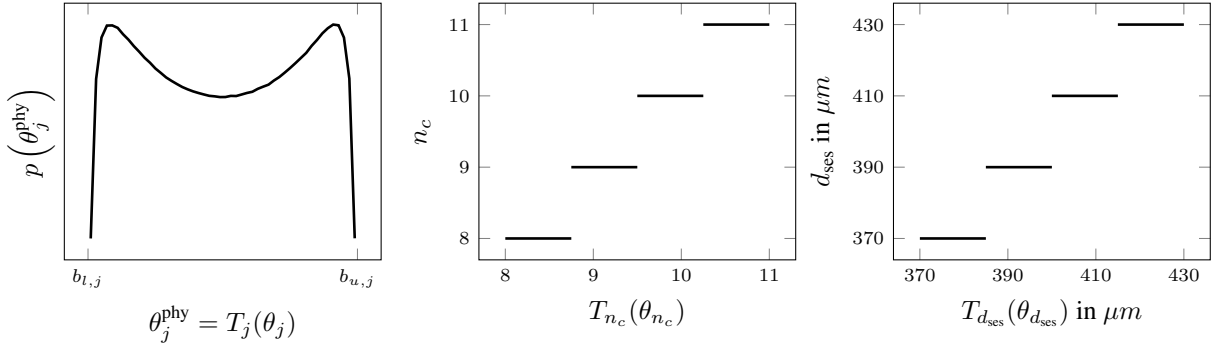
\begin{figure}
    \centering
    \pgfplotsset{
    discard if not/.style 2 args={
        x filter/.code={
            \edef\tempa{\thisrow{#1}}
            \edef\tempb{#2}
            \ifx\tempa\tempb
            \else
                \def\pgfmathresult{inf}
            \fi
        }
    }
}
\pgfplotsset{compat=newest}
\begin{tikzpicture}
  \begin{groupplot}[
    group style={
      group size=3 by 1, 
      horizontal sep=1.3cm, 
    },
    width=\textwidth/2.85,
  ]
   \nextgroupplot[
    ylabel={$\pdf{\lvj^\text{phy}}$},
    xlabel={$\lvj^\text{phy}=T_j(\lvj)$},
    xtick={0,1},
    xticklabels={$b_{l,j}$,$b_{u,j}$},
    ytick=\empty,
    ticklabel style = {font=\scriptsize},
    title style={at={(0.50,0.7)},anchor=north},
  ]
\addplot[color=black,line width=1pt] table [x=x, y=pdf, col sep=comma] {figures/ssb/tanh_pdf.csv};
   \nextgroupplot[
    xlabel={$T_{n_c}(\theta_{n_c})$},
    ylabel={$n_c$},
    xtick={8,9,10,11},
    ytick={8,9,10,11},
    ticklabel style = {font=\scriptsize},
    title style={at={(0.50,0.7)},anchor=north},
  ]
\addplot[color=black,line width=1pt] table [x=c, y=d, col sep=comma,discard if not={b}{0},] {figures/ssb/n_am.csv};
\addplot[color=black,line width=1pt] table [x=c, y=d, col sep=comma,discard if not={b}{1},] {figures/ssb/n_am.csv};
\addplot[color=black,line width=1pt] table [x=c, y=d, col sep=comma,discard if not={b}{2},] {figures/ssb/n_am.csv};
\addplot[color=black,line width=1pt] table [x=c, y=d, col sep=comma,discard if not={b}{3},] {figures/ssb/n_am.csv};

   \nextgroupplot[
    xlabel={$T_{d_\text{ses}}(\theta_{d_\text{ses}})$ in $\mu m$},
    ylabel={$d_\text{ses}$ in $\mu m$},
    xtick={370, 390, 410, 430},
    ytick={370, 390, 410, 430},
    ticklabel style = {font=\scriptsize},
    title style={at={(0.50,0.7)},anchor=north},
  ]
\addplot[color=black,line width=1pt] table [x=c, y=d, col sep=comma,discard if not={b}{0},] {figures/ssb/ses_em.csv};
\addplot[color=black,line width=1pt] table [x=c, y=d, col sep=comma,discard if not={b}{1},] {figures/ssb/ses_em.csv};
\addplot[color=black,line width=1pt] table [x=c, y=d, col sep=comma,discard if not={b}{2},] {figures/ssb/ses_em.csv};
\addplot[color=black,line width=1pt] table [x=c, y=d, col sep=comma,discard if not={b}{3},] {figures/ssb/ses_em.csv};
\end{groupplot}  
\end{tikzpicture}
    \caption{Left: Exemplary prior distribution of a parameter in the physical space. Center: Transformation from continuous to a discrete number of active material particles. Right: Transformation from continuous to discrete separator thickness.}
    \label{fig:parameters}
\end{figure}

Solving the multiphysics nonlinear PDE system for a single parameterization is challenging. This issue is amplified in inverse problems, as certain parameter combinations might lead to diverging simulations. To circumvent this problem, two steps are taken. First, the time step size for the time integration scheme is divided by the C-rate~$c_r$. This couples the numerical discretization to the inverse problem, enabling larger timesteps for smaller C-rates. Secondly, we bound the parameters of interest, see \cref{tab:SSB_parameters}, such that the simulations remain numerically stable. However, to make the inference process unconstrained, the continuous parameters are transformed using:
\begin{align}
    \lvj^\text{phy}=T_j(\lvj)= \frac{1}{2}\left(1+\tanh(\lvj)\right)(b_{u,j}-b_{l,j})+b_{l,j},
\end{align}
here $\boldsymbol{b}_u$ and $\boldsymbol{b}_l$ are the componentwise upper and lower bound of $\lv^\text{phy}$. Due to the manufacturing processes of the composite cells, the active material domain consists of spherical particles. We mimic this behavior by stacking $n_c$ particles in the composite cathode. Similarly, the separator thickness $d_\text{ses}$ is also discrete leading to a variable domain $\Omega=\Omega(n_c, d_\text{ses})$. To avoid the need for discrete random variables, we exploit the black-box nature of the inference schemes by transforming these two quantities by 
\begin{align}
    \lvj^\text{d}=\Delta_j\left\lfloor{\frac{T_j(\lvj)}{\Delta_j}}\right\rfloor\frac{\card{\mathbb{X}_j}}{\card{\mathbb{X}_j}-1} +b_{l,j}
\end{align}
where $\Delta_j=\frac{b_{u,j}-b_{l,j}}{\card{\mathbb{X}_j}}$ is the bin width of the discrete variable. These two transformations allow for unconstrained inference. The prior distribution is set to $\prior = \nd{}{\lv|\boldsymbol{0}, 0.85^2\boldsymbol{I}}$ for all parameters. As shown in \cref{fig:parameters}, this prior choice slightly prefers the bounds of the distribution, which ensures exploration of this parameter space during the initial phase of the inference process.

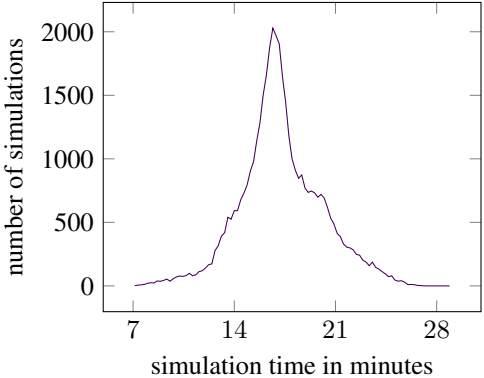
\begin{figure}[h!]
    \centering
    \pgfplotsset{
    discard if not/.style 2 args={
        x filter/.code={
            \edef\tempa{\thisrow{#1}}
            \edef\tempb{#2}
            \ifx\tempa\tempb
            \else
                \def\pgfmathresult{inf}
            \fi
        }
    }
}
\pgfplotsset{compat=newest}
\begin{tikzpicture}
    \definecolor{ABRIS}{RGB}{68, 1, 84}
  \begin{groupplot}[
    group style={
      group size=1 by 1, 
      horizontal sep=2.5cm, 
    },
    colormap/viridis,
    width=0.4\textwidth,
  ]
\nextgroupplot[
    ylabel={number of simulations},
    xlabel={simulation time in minutes},
    xtick = {7,14,21,28},
    ytick = {0,500,1000,1500,2000},
    yticklabels = {0,500,1000,1500,2000}
  ]
\addplot[color=ABRIS, mark size = 2pt] table [x=times, y=pdf, col sep=comma] {figures/ssb/smc/smc_512_46_times.csv};

\end{groupplot}  
\end{tikzpicture}
    \caption{Simulation times for of the 41472 forward model calls made during the SMC512 inference case. Only 4 simulations took longer than 29 minutes, with a maximum of 118 minutes.}
    \label{fig:ssb_sim_times}
\end{figure}

As a reference value, we approximate two reference posterior distributions using a waste-free SMC approach with 512 particles, 10 rejuvenation steps, and a resampling threshold of 0.5. A Metropolis-Hastings kernel was used, requiring 41472 model calls over 7 SMC steps. The evaluation is done batch sequentially with 128 simulations in parallel, leading to a total wall time of approximately 104 hours per reference solution. A histogram of the forward model run times of the solid-state battery model across all SMC steps is shown in \cref{fig:ssb_sim_times}.

This numerical study compares the posterior distribution inferred from ABRIS-BBVI and SMC. For the comparison with SMC, the same setup as the reference simulation, but with 32 particles and 5 rejunevation steps, denoted by the SMC32 case, is chosen. This SMC configuration is chosen in order to reach around 1000 forward model calls for the inference problem.

For the variational case, the variational distribution is constructed as
\begin{align}
    \varq{\lv|\varp} =\sum_{k=1}^{K} \omega_{k}(\varp_{\omega}) \nd{}{\lv|\boldsymbol{\mu} = \varp_{\mu,k},\boldsymbol{C}=\text{diag}(\exp(2\varp_{C,k}))}.
\end{align}
Consequently, the posterior is approximated as a Gaussian mixture model, with $K = 4$ components. The mixture weights are parameterized via a softmax function to ensure unconstrained optimization. In mixture models, the Fisher information matrix for the variational distributions can become singular. This often occurs with the softmax parameterization, where all weights are equal. To handle this in the score estimator condition outlined in \cref{alg:abris_sampling_loop}, we compute the norm of the estimators using the identity matrix $\boldsymbol{A} = \boldsymbol{I}$. For the same reason, no natural gradient, i.e., inverse Fisher matrix preconditioning, of the ELBO gradient is used. In order to mitigate mode collapse and weight degeneracy, $L_2$-regularization term is introduced on the mixture weight parameters $\varp_\omega$, leading to a modified optimization problem 
\begin{align}
\varp^\text{opt}&\in \underset{\varp}{\max} \ \Ex{\varqf}{\ln \unpostf - \ln \varqf} - \alpha_\text{reg} \varp_\omega \cdot \varp_\omega .
\end{align}
The regularization parameter is set to $\alpha_\text{reg}=4 K$, where $K$ is the number of mixture components.
An Adam optimizer with learning rate $0.1$ is used, for the periodic condition, $N_\text{periodic}$ is set to 100. The optimization process is stopped once 1000 forward model calls are reached. The number of samples per batch is set to 4 and the window size $m \in \{20,50\}$. A run for $m=10$ was also done; however, it led to diverging results.

\begin{table}[ht]
    \centering
    \begin{tabular}{c|c|c|c|c|c||c|c}
        Case & $n_\text{sims}$ & $D$ & $c_r$ & $d_\text{ses}$ & $n_\text{c}$ & $\norm{\lv^\text{mode} - \lv^\text{true}}/\norm{\lv^\text{true}}$ & $MMCS$\\
        \hline
        SMC32 & 1312 &  $4.076 \cdot 10^{-14}$ & $0.128$ & $430$ & $9$ & $5.259 \cdot 10^{-6}$ & 0.796\\
        SMC512 & 41472 & $4.018 \cdot 10^{-14}$ & $0.126$ & $430$ & $9$ & $9.573 \cdot 10^{-6}$ & --\\
        \hline
        ABRIS20& 996 & $4.003 \cdot 10^{-14}$ & $0.129$ & $430$ & $9$ & $2.366 \cdot 10^{-6}$ & $\boldsymbol{0.557}$\\
        ABRIS50 & 996 & $4.030 \cdot 10^{-14}$ & $0.129$ & $430$ & $9$ & $\boldsymbol{2.235 \cdot 10^{-6}}$ & 0.946\\
        \hline
    \end{tabular}
    \caption{Comparison of posterior (global) mode, i.e., its difference to the ground truth and maximum marginal Cauchy-Schwarz (MMCS) divergence. For the SMC cases, the mode is selected as the particle with the highest weight; for the ABRIS-BBVI cases, the largest posterior value from 100k samples is selected.}
    \label{tab:ssb_modes}
\end{table}

As can be seen in \cref{tab:ssb_modes}, for most cases, the mode of the posterior approximation is able to be in the vicinity of the ground truth. The difference between the global mode and the true value is the smallest for the ABRIS50 cases, only marginally outperforming the ABRIS20 cases. The SMC cases, the reference case and the case with 32 particles, display a larger difference to the ground truth. Additionally, we compared the posterior distributions to the reference SMC solution using the Cauchy-Schwarz divergence \cite{Principe2000} between posterior marginals, denoted by $MMCS$, via
\begin{align*}
     MMCS(p(\lv), q(\lv)) = \underset{j=1..\dim(\lv)}{\max} -\ln \frac{\int p(\lvj)q(\lvj)d\lvj}{\sqrt{\int p^2(\lvj)d\lvj\int q^2(\lvj)d\lvj}}.
\end{align*}

\begin{figure}
    \centering
    \pgfplotsset{
    discard if not/.style 2 args={
        x filter/.code={
            \edef\tempa{\thisrow{#1}}
            \edef\tempb{#2}
            \ifx\tempa\tempb
            \else
                \def\pgfmathresult{inf}
            \fi
        }
    }
}
\pgfplotsset{compat=newest}
\begin{tikzpicture}
    \definecolor{SMC512}{RGB}{144, 215, 67}
    \definecolor{SMC32}{RGB}{53, 183, 121}
    \definecolor{SMC16}{RGB}{33, 145, 140}
    \definecolor{ABRIS50}{RGB}{49, 104, 142}
    \definecolor{ABRIS20}{RGB}{68, 57, 131}
    \definecolor{ABRIS10}{RGB}{68, 1, 84}
  \begin{groupplot}[
    group style={
      group size=4 by 1, 
      horizontal sep=1cm, 
      vertical sep=1.5cm, 
    },
    width=\textwidth/3.5,
  ]
\nextgroupplot[xlabel={$D\text{ in } \frac{m^2}{s}$},  ticklabel style = {font=\scriptsize}, scaled x ticks=false, xtick={ 1e-14, 4e-14, 1e-13 }, xticklabels={ $10^{-14}$, {\color{red} $D^\text{true}$},$10^{-13}$ }, legend style={at={(0.5,1.01)},legend columns=-1, anchor=south west, draw=none, /tikz/every even column/.append style={column sep=0.75cm}}]
\addplot[color=SMC512, line width=1pt] table [x=diff_am, y=diff_am_pdf, col sep=comma] {figures/ssb/smc/smc_battery_simplified_10_512_seed_46_1d_marginals.csv};\addlegendentry{SMC512};
\addplot[color=SMC32, line width=1pt] table [x=diff_am, y=diff_am_pdf, col sep=comma] {figures/ssb/smc/smc_battery_simplified_5_32_seed_46_1d_marginals.csv};\addlegendentry{SMC32};
\addplot[color=ABRIS20, line width=1pt] table [x=diff_am, y=diff_am_pdf, col sep=comma] {figures/ssb/abris/abris_battery_regcomp_001_4_ncomp_4_seed_46_mem_20_1d_marginals.csv};\addlegendentry{ABRIS20};
\addplot[color=ABRIS50, line width=1pt] table [x=diff_am, y=diff_am_pdf, col sep=comma] {figures/ssb/abris/abris_battery_regcomp_001_4_ncomp_4_seed_46_mem_50_1d_marginals.csv};\addlegendentry{ABRIS50};

\nextgroupplot[xlabel={$c_r\text{ in } -$},  ticklabel style = {font=\scriptsize}, scaled x ticks=false, xtick={ 0.1, 0.13, 0.15 }, xticklabels={ $0.1$,{\color{red} $c_r^\text{true}$},$0.15$ }]
\addplot[color=SMC512, line width=1pt] table [x=c_rate, y=c_rate_pdf, col sep=comma] {figures/ssb/smc/smc_battery_simplified_10_512_seed_46_1d_marginals.csv};
\addplot[color=SMC32, line width=1pt] table [x=c_rate, y=c_rate_pdf, col sep=comma] {figures/ssb/smc/smc_battery_simplified_5_32_seed_46_1d_marginals.csv};
\addplot[color=ABRIS20, line width=1pt] table [x=c_rate, y=c_rate_pdf, col sep=comma] {figures/ssb/abris/abris_battery_regcomp_001_4_ncomp_4_seed_46_mem_20_1d_marginals.csv};
\addplot[color=ABRIS50, line width=1pt] table [x=c_rate, y=c_rate_pdf, col sep=comma] {figures/ssb/abris/abris_battery_regcomp_001_4_ncomp_4_seed_46_mem_50_1d_marginals.csv};

\nextgroupplot[xlabel={$d_\text{ses}\text{ in } \mu m$}, ticklabel style = {font=\scriptsize}, ybar, bar width=0.1cm, xtick={ 370, 390, 410, 430 }, xticklabels={370,390,410,{\color{red} 430}}, ymin=-0.1,ymax=1.1, ytick ={0,0.2,0.4,0.6,0.8,1}]
\addplot[color=SMC512,fill=SMC512] table [x=separator_thickness, y=separator_thickness_pmf, col sep=comma] {figures/ssb/smc/smc_battery_simplified_10_512_seed_46_1d_marginals_separator_thickness_pmf.csv};
\addplot[color=SMC32,fill=SMC32] table [x=separator_thickness, y=separator_thickness_pmf, col sep=comma] {figures/ssb/smc/smc_battery_simplified_5_32_seed_46_1d_marginals_separator_thickness_pmf.csv};
\addplot[color=ABRIS20,fill=ABRIS20] table [x=separator_thickness, y=separator_thickness_pmf, col sep=comma] {figures/ssb/abris/abris_battery_regcomp_001_4_ncomp_4_seed_46_mem_20_1d_marginals_separator_thickness_pmf.csv};
\addplot[color=ABRIS50,fill=ABRIS50] table [x=separator_thickness, y=separator_thickness_pmf, col sep=comma] {figures/ssb/abris/abris_battery_regcomp_001_4_ncomp_4_seed_46_mem_50_1d_marginals_separator_thickness_pmf.csv};

\nextgroupplot[xlabel={$n_c\text{ in } -$}, ticklabel style = {font=\scriptsize}, ybar, bar width=0.1cm, xtick={ 8, 9, 10, 11 }, xticklabels={8,{\color{red} 9},10,11}, ymin=-0.1,ymax=1.1, ytick ={0,0.2,0.4,0.6,0.8,1}]
\addplot[color=SMC512,fill=SMC512] table [x=cathode_thickness_number_of_particles, y=cathode_thickness_number_of_particles_pmf, col sep=comma] {figures/ssb/smc/smc_battery_simplified_10_512_seed_46_1d_marginals_cathode_thickness_number_of_particles_pmf.csv};
\addplot[color=SMC32,fill=SMC32] table [x=cathode_thickness_number_of_particles, y=cathode_thickness_number_of_particles_pmf, col sep=comma] {figures/ssb/smc/smc_battery_simplified_5_32_seed_46_1d_marginals_cathode_thickness_number_of_particles_pmf.csv};
\addplot[color=ABRIS20,fill=ABRIS20] table [x=cathode_thickness_number_of_particles, y=cathode_thickness_number_of_particles_pmf, col sep=comma] {figures/ssb/abris/abris_battery_regcomp_001_4_ncomp_4_seed_46_mem_20_1d_marginals_cathode_thickness_number_of_particles_pmf.csv};
\addplot[color=ABRIS50,fill=ABRIS50] table [x=cathode_thickness_number_of_particles, y=cathode_thickness_number_of_particles_pmf, col sep=comma] {figures/ssb/abris/abris_battery_regcomp_001_4_ncomp_4_seed_46_mem_50_1d_marginals_cathode_thickness_number_of_particles_pmf.csv};
\end{groupplot}  
\end{tikzpicture}
    \caption{Marginal posterior distributions $\pdf{\lvj^\text{phy}|\yobs}$ plotted over the prior sample space for solid-state battery example. The red value indicates the true value for each parameter.}
    \label{fig:ssb_1d_marginals}
\end{figure}
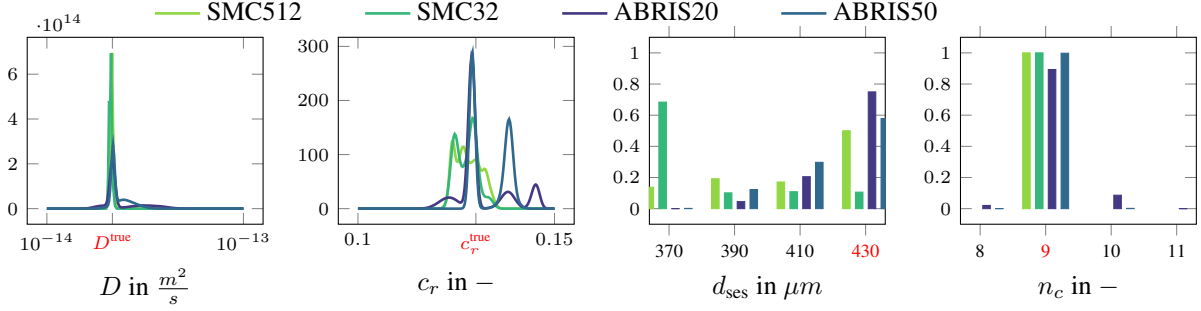

Even though the cases are in agreement regarding the mode, the posterior distributions are different, see \cref{fig:ssb_1d_marginals}. Here, it can be seen that due to the choice of the mixtures and variational objective, the posterior marginals tend to overestimate the uncertainty compared to the reference solution. Particularly for the C-rate marginal, the ABRIS cases place probability mass where the SMC reference solution does not. This phenomenon can also be observed in the number of active material particles marginals, albeit to a much lesser extent. Although the dimensionality of this problem is low, the multimodal nature of the posterior distribution leads to a challenging inference process. This is due to the fact that a cell with a thinner separator thicknesses can show similar discharge curves by an increased C-rate. This ambiguity complicates the inference process, despite the fact that only four parameters need to be inferred. The Cauchy-Schwarz divergence in the $d_\text{ses}$ posterior marginals is 0.05 and 0.07 for the ABRIS cases 50 and 20, respectively, outperforming the SMC32 with 0.80 by an order of magnitude. Hence, the ABRIS algorithm is able to compete with sequential Monte Carlo in these examples. It achieves a higher accuracy, even with approximately 30\% fewer forward model calls. For this particular example, surrogate models could also have been used. However, due to the non-smoothness of the model outputs as observed in \cref{fig:battery_observations}, a large amount of training points would be required to capture the time at which the discontinuity occurs.

\section{Conclusion}
In this work, we extended the black box variational inference (BBVI) approach using importance sampling to reduce the computational cost of the inference process. We introduced a novel sampling procedure by combining meaningful criteria to sample the forward model only when required, leading to a drastic reduction in the computational costs of the inference process, as the forward model is only called when existing evaluations do not provide a sensible gradient estimation. Additionally, the concept of a sampling loop was introduced, enabling adaptive batch-wise sampling within each optimization iteration, which allows for improved Monte Carlo gradient estimation when the user-specific number of samples is insufficient. The resulting inference approach is not limited to a particular family of variational distributions, as long as its density function is differentiable w.r.t. the variational parameters. It is therefore suitable for a wide range of Bayesian inverse problems, without requiring differentiability of the likelihood model. The resulting approach is suitable for Bayesian inference using physics-based simulation models requiring a substantial amount of computational resources. Due to its variational nature, the achievable accuracy depends on the choice of variational distribution family. Therefore, the adaptive batch-sized regulated importance sampling (ABRIS) algorithm is particularly well-suited for variational inference where the posterior distribution can be well approximated by simple posteriors, for continuous and discrete parameters, even in moderately high-dimensional problems. This is reflected in the first two numerical studies, where the ABRIS enhancement outperforms plain BBVI, SMC, and MCMC by orders of magnitude in terms of model calls, without the need to forward model gradients or surrogates. For more complex posterior distributions, i.e., the solid-state battery example, a performance gain can still be observed, although to a smaller extent. Here, the benefit can be increased by changing the variational objective to tighter variational bounds \cite{Zhang2017} tailored to the complex distribution itself.

\FloatBarrier
\section*{Acknowledgements}
GRR, CPS and WAW acknowledge support from the German Federal Ministry of Education and Research (project FestBatt2, \texttt{03XP0435B}). MD, JN and WAW acknowledge support by BREATHE, an \texttt{ERC-2020-ADG} Project, Grant Agreement ID \texttt{101021526}. 

\appendix

\section{Solid-state battery modeling }\label{sec:appendix_SSB}
We briefly list the boundary, interface, and initial conditions. The mass flux density is denoted by $\boldsymbol{j}$, and $k$ is the equivalent spring stiffness of the battery housing. For in-depth information, the reader is referred to \cite{Schmidt2023}.

\begin{align}
\boldsymbol{i}\cdot \boldsymbol{n}=0 \text{ on } \Gamma_\text{cut} \times T\\
\boldsymbol{j}\cdot \boldsymbol{n}=0 \text{ on } \Gamma_\text{cut}\cup \Gamma_\text{cc,a-o}\cup\Gamma_\text{cc,c-o} \times T\\
\Phi=0 \text{ on } \Gamma_\text{cc,a-o} \times T\\
\boldsymbol{j}_\text{ed} \cdot \boldsymbol{n}_\text{ed,el} = \frac{1}{F}\boldsymbol{i}_\text{ed} \cdot \boldsymbol{n}_\text{ed,el} = \frac{1}{F} \boldsymbol{i}_\text{el} \cdot \boldsymbol{n}_\text{ed,el} = \boldsymbol{j}_\text{el} \cdot \boldsymbol{n}_\text{ed,el} \text{ on } \Gamma_\text{ed-el} \times T\\
\boldsymbol{u}_\text{def}=\boldsymbol{0} \text{ on } \Gamma_\text{cc,c-o} \times T \\ 
(\boldsymbol{F}\cdot\boldsymbol{S})\cdot \boldsymbol{n}=k(\boldsymbol{u}\cdot\boldsymbol{n}-u_\text{cc,a-o}) \text{ on } \Gamma_\text{cc,a-o} \times T \\ 
\boldsymbol{u}_\text{def}\cdot \boldsymbol{n}=0 \text{ on } \Gamma_\text{cut} \times T \\ 
c=c^0_\text{ed} \text{ in } \Omega_\text{ed} \times \{0\}\\
c=c^0_\text{el} \text{ in } \Omega_\text{el} \times \{0\}\\
\boldsymbol{u}=\boldsymbol{0} \text{ in } \Omega \times \{0\} \\ 
\dot{\boldsymbol{u}}=\boldsymbol{0} \text{ in } \Omega \times \{0\}
\end{align}
\printbibliography
\end{document}